\documentclass[10pt, journal, journal]{IEEEtran}
\usepackage{bbding}
\usepackage{pifont}
\usepackage{wasysym}
\usepackage{amssymb}

\usepackage{multirow}

\usepackage{tikz}
\usetikzlibrary{arrows.meta,positioning,calc,fit,backgrounds}

\usepackage{cite}
\usepackage{tabularx}        
\usepackage{booktabs}        
\usepackage{makecell}        
\usepackage{threeparttablex} 
\usepackage{graphicx}
\usepackage{xurl}   
\usepackage{hyperref}
\hypersetup{hidelinks}
\ifCLASSINFOpdf
\else
\fi
\usepackage{url}

\begin{document}
%
\title{From Network Automation to Trustworthy Autonomous Networking in the LLM Era: A Network Control Intelligence Perspective}
%
%
%

\author{Tianzhu~Zhang, 
        Changgang~Zheng, 
        Shanshan~Wang,
        Yarui~Zhang,
        Lina~Shi,
        Yue~Jin,
        Xiaofei~Wang,
        Meikang~Qiu
\thanks{
T. Zhang, L. Shi, and Y. Jin are with Nokia Bell Labs, Massy, France. (Email: tianzhu.zhang@nokia-bell-labs.com, lina.1.shi@nokia.com, yue.1.jin@nokia-bell-labs.com)
}
\thanks{C. Zheng is with the University of Oxford, Oxford, UK. (Email: changgang.zheng@eng.oxfordalumni.org)
}
\thanks{S. Wang is with Telecom Paris, Institut Polytechnique de Paris, Palaiseau, France. (Email: shanshan.wang@telecom-paris.fr)}
\thanks{Y. Zhang is with École Normale Supérieure Paris-Saclay, Université Paris-Saclay, Gif-sur-Yvette, France. (Email: yarui.zhang@ens-paris-saclay.fr)}
\thanks{X. Wang is with Tianjin University, Tianjin, China. (Email: xiaofeiwang@tju.edu.cn)}
\thanks{M. Qiu is with Augusta University, Augusta, Georgia, USA. (Email: qiumeikang@gmail.com)
}
}

\maketitle

\begin{abstract}
Since the inception of modern communication networks, the quest for operations automation has never ceased. Yet the evolution of network automation is difficult to characterize with a single maturity ladder. Throughout this history, network control systems have expanded their capabilities for observation, decision support, routine execution, and operator interaction, but these capabilities have not advanced uniformly. Such uneven progress makes the degree of automation an unreliable proxy for trustworthy network-side actuation. The unresolved question is not simply how much automation a system provides, but under what conditions it can be entrusted to change the network state. 
This paper examines that question through Network Control Intelligence (NCI), a five-axis framework spanning Decision Logic, Adaptability, Knowledge, Control Delegation, and Interface. We use NCI to organize the evolution of network-control systems into three eras: rule-based and scripted automation, programmable and data-driven control, and Large Language Model (LLM)-enabled network operations. 
Viewed through this framework, the three eras reveal a persistent asymmetry. Earlier automation made monitoring and routine control repeatable; programmable and data-driven control expanded optimization, telemetry, and domain-specific closed loops; and LLM-enabled operations now strengthen intent interpretation, evidence synthesis, change authoring, and workflow coordination. None of these gains, however, automatically determines when network control should be trusted to change the network state. We frame trustworthy autonomy as a governed alignment between what a system can infer, what it can verify, and what it is authorized to execute. On that basis, the paper develops a reference architecture that separates proposal generation from governed execution, identifies recurring integration patterns for LLM-enabled operations, and derives a research agenda for higher network autonomy under explicit assurance, safety, and governance constraints.
\end{abstract}


%

\IEEEpeerreviewmaketitle


\section{Introduction}
\label{sec:intro}
Network operations have long been driven by a persistent ambition: to reduce human intervention while maintaining network reliability and service assurance~\cite{schoenwaelder2003overview}. Over decades, this movement has reshaped network infrastructure from hardware-centric architectures built around dedicated network appliances toward software-defined, virtualized, and programmable systems~\cite{kreutz2014software,isg2013network,mckeown2008openflow,bosshart2014p4}. In parallel, network operations and management have shifted from rule-based, operator-driven workflows toward programmable control loops informed by telemetry and data-driven analytics~\cite{boutaba2018comprehensive,ayoubi2018machine}. Beyond these technical shifts, the same operational ambition has been framed through several related visions, including autonomic networking~\cite{kephart2003vision}, self-driving networks~\cite{feamster2017and}, zero-touch network and service management (ZSM)~\cite{etsi-gs-zsm}, and autonomous networking~\cite{tmforum-regional-an-progress}.
These visions share a practical concern: how much intelligence and authority can be moved into the control loop without compromising network reliability, accountability, or service quality?
This paper uses the term ``autonomous networking" to refer to network control systems that perceive the operational state, reason over objectives and constraints, and recommend or execute actions within a governed operational envelope.

With the global rollout of 5G and the expected progression toward 6G, network systems are expected to face greater scale, heterogeneity, and service dynamism~\cite{shen2023five}. This growing operational complexity has elevated autonomous networking from a long-term research vision into a strategic priority for many network operators~\cite{capegemini-report}.
At the technical level, however, current automation techniques can boost operational efficiency but remain largely constrained in their ability to interpret operator intent, synthesize heterogeneous operational evidence, and coordinate decisions across domains and timescales~\cite{etsi-zsm-016,etsi-zsm-009, IG1251E, IG1414}. These limitations suggest that network autonomy cannot be judged solely by the sheer number of operations automated. It also depends on how decisions are formed, how behavior adapts, what operational knowledge is available, where authority is placed, and how objectives and evidence are exchanged. We refer to this set of capabilities as Network Control Intelligence (NCI), the structured control capability through which network systems transform observations, objectives, policies, and operational knowledge into recommendations or actions under explicit authority, safety, and auditability constraints~\cite{wang2024netassistant,wang2025intent,mekrache2025oss,wang2024netconfeval,wei2025inta,ahmed2026vision,he2025just}. 

\begin{figure*}[!tb]
\begin{center}
\includegraphics[width=\textwidth]{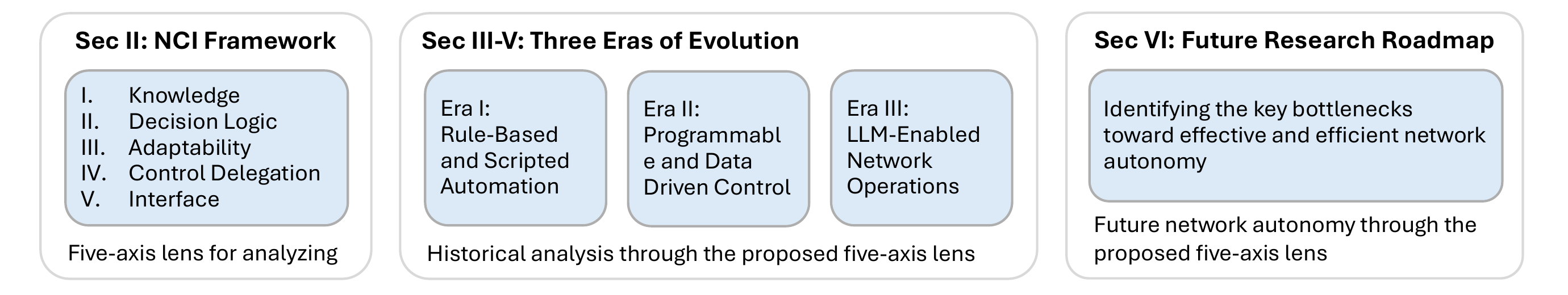}
\caption{Organization of the paper. The paper first introduces the Network Control Intelligence (NCI) framework, then uses its five axes to analyze three stages in the evolution of autonomous networking, and finally derives a roadmap for trustworthy progress toward higher levels of autonomy.}
\label{fig:org}
\end{center}
\end{figure*}

In this paper, we use NCI as a profiling lens to compare network control systems in terms of their underlying control properties, including how they make decisions, adapt over time, represent and use operational knowledge, delegate control authority, and exchange objectives, evidence, and feedback.
The five-axis view also gives the historical development of autonomous networking a more precise structure. The development of NCI unfolded through successive changes in how networks were observed, reasoned about, programmed, and governed. In its early days, network operations and management, often termed Operations, Administration, and Maintenance (OAM), relied on human operators working through Operations Support Systems (OSS) and Network Management Systems (NMS), via static procedures, scripts, and control logic specified by domain experts~\cite{bell1983engineering}. These approaches made routine work more repeatable, but their rule sets and workflows remained limited by what human teams could engineer, audit, and maintain. As topologies, service portfolios, and traffic patterns became more diverse, handcrafted control logic struggled to keep pace with operational complexity. This limitation motivated the evolution from scripted procedures toward programmable control. The subsequent wave of network softwarization and programmability gave the long-standing automation agenda a stronger technical foundation. Logically centralized control, virtualized services, extensive telemetry, and data-driven analytics enabled finer-grained traffic steering, elastic service composition, and policy-driven orchestration~\cite{boutaba2018comprehensive}. This shifted network operations from script-based management to domain-specific optimization, e.g., traffic engineering, energy management, and Radio Access Network (RAN) resource allocation~\cite{ayoubi2018machine}. Nonetheless, many resulting control loops remained domain-specific, with limited support for end-to-end decision-making aligned with operator intent~\cite{huang2024large}. Human oversight still plays a central role in orchestrating complex workflows, handling unanticipated situations, and reconciling competing objectives.

The rise of large language models (LLMs) adds another chapter to this history. Unlike conventional optimizers or domain-specific controllers, LLM-enabled systems are well suited to tasks that require intent interpretation, evidence synthesis, and multi-step workflow composition. When integrated with retrieval systems, external tools, and operational interfaces, they can interpret requests using relevant operational records, invoke appropriate tools, and assemble workflows~\cite{toolllm,yao2023react}. In network operations, they can generate structured, auditable outputs, such as diagnostic interpretations~\cite{wang2024netassistant}, operational procedures~\cite{wang2025intent}, and candidate configuration changes~\cite{wei2025inta}. These outputs can improve human- or controller-mediated decision-making, but they do not remove the need for governed execution.
In practice, LLM-enabled operational systems are seldom isolated prompt-response chatbots. They are increasingly embedded in agentic workflows that retrieve operational context, invoke external tools, maintain state across steps, validate intermediate outputs, and recover from failed actions. In such settings, long-horizon reliability depends not only on model capability, but also on how these workflow mechanisms are designed, regulated, and evaluated~\cite{anthropic2025context,anthropic2026evals,anthropic2025harnesses,openai2026harness}. In network operations, such agents can improve semantic mediation across operational interfaces and support workflows aligned with operator policies and Service-Level Objectives (SLOs), while execution remains mediated by domain controllers, verifiers, authorization mechanisms, rollback procedures, and change-management systems~\cite{wang2024netassistant,wang2025intent,mekrache2025oss}. Recent LLM-enabled network-operation systems and industry prototypes point to an incremental architecture: LLMs help assemble intent, evidence, and candidate workflows, while authoritative control remains anchored in specialized control loops and governed execution paths~\cite{xiao2025sannet,blueplanet,huawei-agent,netcracker,deutsche}.


This progression motivates the three-era organization used in this paper. The eras are not separated by chronology alone, but by changes in how NCI is realized and governed. Rule-based and scripted automation refers to systems in which operational intelligence is primarily encoded in human-defined procedures, scripts, alarms, and management rules. Programmable and data-driven control refers to systems in which software-defined control, virtualization, telemetry, optimization, and learning shift decision-making to programmable controllers and analytics pipelines. LLM-enabled network operations refer to systems in which language-model components assist with intent interpretation, evidence synthesis, and workflow composition, while authoritative execution remains outside the model.
Although earlier mechanisms often persist even after newer ones emerge, the three-era structure captures successive shifts in how NCI is expressed and constrained.

Within each era, this paper reviews representative network-control systems. The selected systems satisfy one or more of four criteria: they introduce or exemplify a mechanism that changes at least one NCI axis; they have influenced research, standards, or operational practice; they provide sufficient public technical detail to support axis-level profiling; or they illustrate a distinct role in the control stack, such as monitoring, configuration, traffic engineering, fault diagnosis, orchestration, or LLM-enabled workflow composition. This selection strategy allows the paper to show how NCI has evolved across its five axes and why progress toward autonomy has remained uneven.

The main contributions of this paper are as follows:
\begin{itemize}
\item It introduces the Network Control Intelligence (NCI) framework for profiling network control systems across five coupled axes: Decision Logic, Adaptability, Knowledge, Control Delegation, and Interface.

\item It applies this framework to representative network-control systems across three eras, selected for their influence, technical specificity, and ability to illustrate shifts along the NCI axes, showing that the evolution of network-control intelligence has been uneven and is not reducible to a single maturity trajectory.

\item It characterizes the LLM-enabled era by separating proposal generation from governed execution, in which LLM-enabled components improve intent interpretation, evidence synthesis, change authoring, and workflow coordination, while execution remains mediated by controllers, verification, authorization, and change-management workflows.

\item It derives a forward-looking roadmap for trustworthy autonomous networking, showing how future progress depends on the concerted maturation of operational knowledge, decision logic, adaptability, the Control Delegation axis, and operator interfaces under explicit assurance constraints.
\end{itemize}


As illustrated in Fig.~\ref{fig:org}, the remainder of this paper is organized as follows. Section~\ref{sec:framework} introduces the NCI framework and situates it relative to standards, industry frameworks, and related literature. Sections~\ref{sec:era-I}--\ref{sec:era-III} trace the evolution of autonomous networking across three eras through the five-axis NCI lens. Section~\ref{sec:challenges} develops the forward-looking roadmap for trustworthy progress toward higher levels of network autonomy. Section~\ref{sec:conclusion} concludes the paper.


\begin{figure*}[!tb]
\centering
\includegraphics[width=1\textwidth]{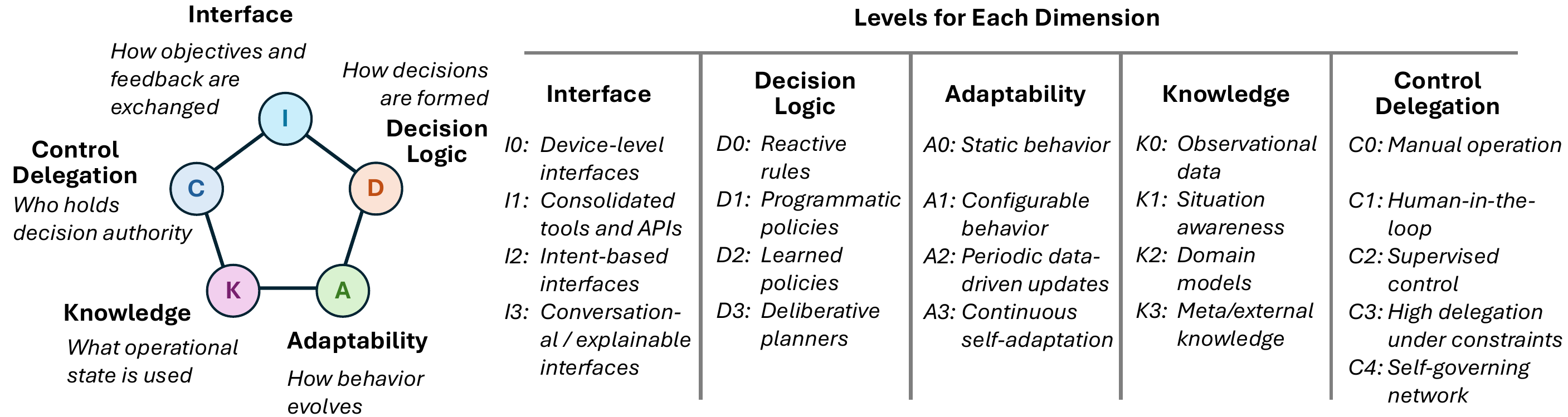}
\caption{The five-axis framework for Network Control Intelligence. The axes describe complementary control properties: how decisions are formed, how behavior adapts, what operational knowledge is available, how authority is delegated, and how objectives, evidence, and feedback are exchanged. In a control loop, observations are interpreted through operational knowledge and refined through interfaces. Decision logic and adaptation use this knowledge to produce recommendations or actions, while Control Delegation defines whether a system’s outputs remain advisory, require human approval, or can be executed automatically within an authorized scope.}
\label{fig:fiveaxis}
\end{figure*}

\section{Network Control Intelligence for Autonomous Networking}
\label{sec:framework}

To characterize network control systems, we use an abstract control model. In this model, a network is represented as a collection of domains, such as the RAN, transport, core, and data center or cloud/edge domains. Each domain provides telemetry channels that expose \emph{observables}, including counters, events, key performance indicators (KPIs), logs, traces, alarms, and topology state. It also provides control and management interfaces that expose \emph{actuators}, such as configuration knobs, application programming interfaces (APIs), and orchestration hooks.
Within this model, an \emph{NCI component} denotes any control or management function that consumes observables, interprets them in light of explicit objectives, policies, and operational knowledge, and produces recommendations or actions for the relevant actuators. Depending on its authorized role, the component may assist human operators, propose candidate changes, execute scoped control actions, or participate in end-to-end orchestration. NCI profiles the role that such a function plays in a control process; it is not an intrinsic intelligence score for a software artifact.

From a control perspective, autonomous networking can be viewed as the gradual enrichment of the classical sense-think-act loop. In this paper, \emph{open-loop automation} denotes predefined execution without runtime feedback correction. \emph{Closed-loop automation} denotes feedback-based operation in which monitored outcomes can trigger corrective actions. \emph{Control-loop autonomy} further implies that feedback is interpreted against explicit objectives, constraints, and predictive or adaptive models within an authorized operational envelope. The network is sensed through telemetry, events, topology, inventory, and service-state observations; these observations are interpreted using operational knowledge and decision logic; and candidate actions are applied via controllers, orchestrators, or management interfaces, subject to policy and assurance constraints. NCI extends this control view by making five properties explicit: how decisions are formed, how behavior adapts, how operational knowledge is represented, how control authority is delegated, and how objectives and evidence are exchanged, as illustrated in Fig.~\ref{fig:fiveaxis}.

\subsection{Network Control Intelligence: a five-axis view}
The NCI framework consists of five analytically distinct but operationally coupled axes: \textbf{Decision Logic}, \textbf{Adaptability}, \textbf{Knowledge}, \textbf{Control Delegation}, and \textbf{Interface}. Each axis captures a structural dimension of an autonomous network system, as shown in Fig.~\ref{fig:fiveaxis}. The framework classifies systems by evolvable control properties, regardless of implementation technology. Legacy scripted tools, optimization-based controllers, learning-driven systems, and LLM-enabled agents can thus all be described within the same five-dimensional space. This makes it possible to compare, in a common vocabulary, how different systems form decisions, adapt to changing conditions, represent operational knowledge, delegate control authority, and interact with external actors.

\subsubsection*{Framework Design Rationale}
\label{sec:framework-rationale}

The five-axis framework recasts the control-loop view as a set of profiling dimensions. The axes do not correspond to separate implementation modules; instead, they describe how decisions are formed, how behavior changes over time, how operational knowledge is represented, how control authority is delegated, and how objectives and feedback are exchanged. This distinction is important because two systems may use similar controller architectures yet differ sharply in delegation, knowledge scope, adaptability, or interface abstraction. Conversely, systems built from different technologies may occupy similar positions in the NCI space if they expose comparable control properties.

The choice of these five axes follows from the minimal functional path of a network-control process. Classical feedback and cybernetic views emphasize sensing, state, feedback, and action in regulated systems~\cite{ashby1956introduction,aastrom2021feedback}, while agent-oriented models describe intelligent behavior as the mapping from perceptual input and objectives to action~\cite{russell2021artificial}. In operational settings, this mapping also depends on situation awareness~\cite{endsley2017toward}, the system's ability to adapt to changing conditions~\cite{sutton1998reinforcement}, the allocation of authority between humans and automation~\cite{parasuraman2000model}, and the interfaces through which objectives, evidence, and feedback are exchanged. These considerations motivate the five NCI axes: \emph{Knowledge}, \emph{Decision Logic}, \emph{Adaptability}, \emph{Control Delegation}, and \emph{Interface}. Other important concerns, including security, assurance, governance, scalability, interoperability, explainability, cost, and performance, are treated as cross-cutting constraints rather than additional axes. They determine whether movement along the five axes is safe, reliable, and operationally defensible, but they do not describe a separate functional step in the control loop~\cite{leveson2016engineering}.

The purpose of the framework is diagnostic rather than to assign a single ordered autonomy level. It asks which control properties are present, which are missing, and which assumptions must hold for a given NCI profile to be operationally credible. This makes it possible to compare systems without forcing them onto a single autonomy ladder. The axes are analytically distinct but not independent; richer knowledge can change what decision logic can verify, stronger interfaces can change what operators can safely approve, and broader autonomy can expose weaknesses in adaptation or assurance.

\subsubsection*{Decision Logic}
\label{sec:axis-decision}

The \textbf{Decision Logic} axis describes how decisions are formed. It characterizes how an NCI component maps observations, objectives, and constraints to recommendations or actions, ranging from fixed reactive rules to explicit optimization, learned policies, and deliberative planning. This axis should not be read as a simple quality ladder: optimization-based controllers can be more auditable and constraint-aware than learned policies for some tasks. In contrast, learned policies may offer faster inference or better empirical generalization in others. For profiling, we distinguish four representative rungs:
\begin{itemize}
    \item \textbf{D0 (Reactive Rules):} Hand-crafted conditions, thresholds, or event-condition-action rules that map current observations to immediate responses.
    \item \textbf{D1 (Programmatic Policies):} Structured control programs, policy engines, or optimization mechanisms operating over explicit objectives, models, or constraint sets.
    \item \textbf{D2 (Learned Policies):} Data-driven mappings from observations to actions or recommendations, trained from historical, simulated, or online experience.
    \item \textbf{D3 (Deliberative Planners):} Goal- and plan-based controllers that evaluate alternatives, reason over future network states, and revise plans as conditions evolve.
\end{itemize}
Note that D3 systems may integrate D1-style solvers and D2-style learned components into a broader planning loop; the defining property is the ability to reason about alternatives and multi-step consequences.

\subsubsection*{Adaptability}
\label{sec:axis-adapt}

This axis describes how behavior changes over time. It captures how an NCI component changes its behavior as network conditions, traffic patterns, objectives, or policies evolve. It distinguishes systems whose behavior is fixed at design time from systems whose parameters, models, or policies are updated periodically or continuously. For profiling, we distinguish four regimes:
\begin{itemize}
    \item \textbf{A0 (Static Behavior):} Fixed rules, policies, or parameters; any change requires manual reconfiguration.
    \item \textbf{A1 (Configurable Behavior):} Predefined modes, profiles, or parameters that operators can select or tune, possibly aided by basic analytics.
    \item \textbf{A2 (Periodic Data-Driven Updates):} Models, parameters, or configurations are periodically re-optimized from accumulated data, typically under human oversight.
    \item \textbf{A3 (Continuous Self-Adaptation):} The NCI detects drift and updates its models, policies, or strategies online within an authorized operational envelope.
\end{itemize}
Higher adaptability can improve robustness under non-stationarity, but it also increases the need for monitoring, regression control, and safety constraints to prevent unreliable updates.
Adaptability also depends on what state and context the component can observe or represent.

\subsubsection*{Knowledge}
\label{sec:axis-knowledge}

The \textbf{Knowledge} axis describes what operational context is available to an NCI component and how that context is structured. It includes raw telemetry, inferred state, topology, and service models, policies, constraints, external documentation, and higher-order knowledge about model scope or uncertainty. For profiling, we distinguish four layers:
\begin{itemize}
    \item \textbf{K0 (Observational Data):} Raw or lightly processed measurements, such as counters, system logs, events, flow records, and link metrics.
    \item \textbf{K1 (Situation Awareness):} Structured views of current network state, such as topology, active traffic matrices, detected anomalies, alarms, and service-impact context.
    \item \textbf{K2 (Domain Models):} Predictive, causal, policy, or simulation models that represent how a domain behaves and how candidate actions may affect it.
    \item \textbf{K3 (Meta and External Knowledge):} Cross-domain or external knowledge sources, such as knowledge graphs, digital twins, documentation, historical incident records, and explicit knowledge about model limitations, uncertainty, or freshness.
\end{itemize}
The Knowledge axis is not only about data volume. Its central concern is whether the NCI has a faithful, structured, and up-to-date representation of the operational context required for safe decision-making.


\subsubsection*{Control Delegation}
\label{sec:axis-control-delegation}

This axis describes where control authority resides. It captures how decision authority is delegated between human operators and automated components. It is distinct from decision quality: a system may have sophisticated reasoning but low delegation if its outputs remain advisory, whereas a simpler controller may have higher delegation within a narrow, well-verified domain. For profiling, we use five delegation rungs:
\begin{itemize}
    \item \textbf{C0 (Manual Operation):} The system provides information or scripted assistance, but humans select and approve all substantive actions.
    \item \textbf{C1 (Human-in-the-Loop):} The system can propose actions with explicit human approval for execution.
    \item \textbf{C2 (Supervised Control):} The system autonomously handles routine tasks within predefined guardrails and escalates unusual, ambiguous, or high-risk cases.
    \item \textbf{C3 (High Delegation Under Constraints):} The system makes and executes decisions across a broader scope without per-action approval, while operating within explicit policy, safety, and escalation constraints.
    \item \textbf{C4 (Self-Governing Network):} The system is self-governing within a declared operational design domain, planning, executing, and adapting strategies under high-level human intent and explicit assurance assumptions.
\end{itemize}
Note that higher delegation must be justified by capabilities on the other axes and by the surrounding assurance envelope. In practice, increases in delegation require stronger evidence that decision logic, adaptability, and operational knowledge are adequate for the authorized scope. Verification, rollback, and escalation mechanisms help bound the associated risk, but they cannot substitute for inadequate control capability or operational knowledge.

\begin{table*}[t]
\centering
\scriptsize
\caption{How major standards and industry frameworks relate to the five-axis NCI framework. Cells indicate whether an initiative emphasizes an axis (Strong), covers it in a limited or domain-scoped manner (Partial), assumes it indirectly (Implicit), or largely treats it as out of scope (--). The mapping is interpretive: it characterizes the main emphasis of each initiative without assigning formal compliance levels.}
\label{tab:standards-vs-nci}
\begin{tabular}{p{3.2cm} p{5.1cm} c c c c c}
\toprule
\textbf{Initiative} &
\textbf{Primary emphasis} &
\textbf{Decision Logic} &
\textbf{Adaptability} &
\textbf{Knowledge} &
\textbf{Control Delegation} &
\textbf{Interface} \\
\midrule

TM Forum~\cite{ig1392,anm,IG1414} &
Maturity levels and operational transformation &
Implicit & Implicit & Partial & Strong & Partial \\

ETSI ZSM~\cite{etsi-gs-zsm} &
Closed-loop management across domains &
Partial & Partial & Partial & Partial & Strong \\

ETSI ENI~\cite{eni} &
Cognitive management and context-aware policy &
Partial & Partial & Strong & Partial & Partial \\

3GPP SON~\cite{ts32500} &
RAN automation loops &
Partial & Partial & Partial & Partial & Partial \\

O-RAN~\cite{o-ran} &
RAN automation platform with programmable control apps and management interfaces &
Partial & Partial & Partial & Partial & Strong \\

3GPP NWDAF~\cite{3gpp.29.520} &
Standardized analytics services for 5G core &
Implicit & Implicit & Strong & -- & Strong \\

ITU-T FG-AN~\cite{fg-an} &
Pre-standardization guidance for autonomous-network evolution &
Implicit & Implicit & Partial & Partial & Implicit \\

ITU-T Y.317x~\cite{itu-3172} &
Reference architecture for ML in future networks &
Partial & Partial & Strong & Implicit & Partial \\

NETCONF/RESTCONF/YANG, OpenConfig/gNMI~\cite{enns2011network,bierman2017restconf,bjorklund2016yang,claise2019network,openconfig} &
Management protocols, data models, and telemetry interfaces &
-- & -- & Partial & -- & Strong \\

IETF intent RFCs~\cite{clemm2022intent,li2022rfc} &
IBN concepts and intent classification methodology &
Partial & Implicit & Partial & Implicit & Strong \\

IETF ANIMA ACP, BRSKI~\cite{eckert2021rfc,pritikin2021bootstrapping} &
Secure the autonomic control plane and bootstrap the infrastructure &
Implicit & Implicit & Partial & Partial & Strong \\

OpenFlow/ONOS~\cite{mckeown2008openflow,onos} &
SDN programmability and controller platforms &
Partial & Partial & Partial & Partial & Strong \\

ONF CORD/SEBA~\cite{cord,das2021cord} &
Reference designs for virtualized access/edge &
Partial & Implicit & Partial & Partial & Strong \\

\bottomrule
\end{tabular}
\end{table*}

\subsubsection*{Interface}
\label{sec:axis-interface}

The \textbf{Interface} axis describes how objectives, constraints, commands, state, explanations, and feedback are exchanged between the NCI and external actors, including operators, controllers, management systems, and peer automation components. It captures the abstraction level of interaction, from device-level commands to intent-based and conversational interfaces. For profiling, we distinguish four levels:
\begin{itemize}
    \item \textbf{I0 (Device-Level Interfaces):} Low-level command-line interfaces, device-specific configuration scripts, or narrow management interfaces.
    \item \textbf{I1 (Consolidated Tools and APIs):} Centralized management tools, controllers, dashboards, and structured APIs that aggregate state and expose programmable operations.
    \item \textbf{I2 (Intent-Based Interfaces):} Policy- or intent-level specifications that describe desired outcomes, constraints, or service objectives instead of concrete commands.
    \item \textbf{I3 (Conversational and Explainable Interfaces):} Natural-language, multimodal, or dialogue-based interfaces that support clarification, explanation, and human-facing supervision of autonomous behavior.
\end{itemize}
As interfaces move upward, they reduce the burden of translating high-level objectives into low-level commands, but they also increase the need for precise intent semantics, evidence-bearing explanations, and machine-checkable constraints. Higher interface abstraction is not always more operationally mature: a machine-checkable intent interface may be safer than a conversational interface when ambiguity, authorization, or verification is difficult to resolve.

The framework is most useful when applied comparatively. For example, an SNMP-based alarm system is typically placed near \emph{D0/A0/K0--K1/C0--C1/I1}: it reacts through fixed thresholds, relies on narrow observational state, and usually requires human interpretation and approval. A traffic-engineering controller such as Google's B4 is closer to \emph{D1/A1/K2/C2/I1}: it optimizes over explicit objectives and domain models and can execute within a scoped authority envelope. In comparison, an LLM-based change-authoring assistant may reach \emph{D2/K3/I3} in proposal generation and evidence synthesis while remaining at only \emph{C0--C1}, because commit authority remains outside the model. These profiles clarify why network-control systems that all appear ``intelligent'' can differ sharply in operational autonomy. Stronger observation, reasoning, or interface support should not be equated with broader control authority.


\subsection{Standards Landscape}
\label{sec:standards-landscape}

The NCI framework is not intended to replace existing autonomous-network standards, maturity models, or management architectures. It provides a finer-grained analytical vocabulary for comparing the control-intelligence properties that these initiatives only partially emphasize. Standardization bodies and industry fora have developed several related frameworks for autonomous, self-driving, zero-touch, or intent-based networking, but their purposes differ. Some define maturity levels for operational transformation, some specify closed-loop management architectures, some standardize domain-specific analytics functions, and others provide protocols and data models for network automation. Table~\ref{tab:standards-vs-nci} summarizes how these initiatives relate to the five NCI axes.

The first group focuses on how far operational authority can be delegated to automated systems. TM Forum's Autonomous Networks program defines a staged Level~0--Level~5 model for assessing the degree of autonomous operation~\cite{ig1392,anm,IG1414}. From the NCI perspective, such maturity levels provide a coarse-grained view of the Control Delegation axis. They indicate the extent to which operational authority is transferred from humans to automated systems. Regional progress material further shows why a single scalar level is often insufficient. Different operators, domains, and operational processes may advance at varying rates, with service assurance, fulfillment, transport, RAN, core, and service delivery occupying distinct maturity levels~\cite {tmforum-regional-an-progress}. The remaining NCI axes help characterize how a given maturity level is realized in practice, for example, through particular forms of decision logic, knowledge representation, adaptability, or interface abstraction. ITU-T activities on autonomous networks and intelligence levels play a related role by framing autonomy and machine-learning integration at a cross-technology level~\cite{fg-an,itu-3172}.

The second group concentrates on closed-loop management and domain-specific automation. ETSI ZSM and ENI define architectural patterns for zero-touch and cognitive management, including management domains, data fabrics, assurance functions, and policy-driven closed loops~\cite{etsi-gs-zsm,eni}. 3GPP SON, NWDAF, and O-RAN provide more domain-specific mechanisms for RAN and 5G core automation, including self-configuration, self-optimization, standardized analytics exposure, and RIC-based control applications~\cite{ts32500,3gpp.29.520,o-ran}. In NCI terms, these initiatives cover selected parts of Decision Logic, Adaptability, Knowledge, Control Delegation, and Interface. They specify important architectural and domain-specific mechanisms, but they do not provide a general framework for comparing systems across all five axes.

The third group foregrounds \emph{interfaces, protocols, and programmability}. The IETF/IRTF works on NETCONF~\cite{enns2011network}, RESTCONF~\cite{bierman2017restconf}, YANG~\cite{bjorklund2016yang}, intent-based networking~\cite{clemm2022intent,li2022rfc}, and autonomic networking~\cite{eckert2021rfc,pritikin2021bootstrapping} clarify how network state, configuration, and intent can be represented and exchanged. Operational practice around OpenConfig/gNMI similarly strengthens structured configuration and telemetry exchange~\cite{claise2019network,openconfig}. ONF efforts, including OpenFlow~\cite{mckeown2008openflow}, ONOS~\cite{onos}, CORD~\cite{cord}, and SEBA~\cite{das2021cord}, provide concrete examples of controller-centered programmability and open operational interfaces. These efforts primarily strengthen the Interface axis and, in controller-centered settings, parts of Decision Logic and Knowledge.

Taken together, these initiatives point toward network operations that are more programmable, more closed-loop, more intent-aware, and more autonomous within scoped authority. Their emphasis, however, remains uneven across the five NCI axes. A system may be described as highly autonomous in a maturity model while still relying on narrow knowledge, static adaptation, or low-level interfaces. Conversely, a system may expose rich intent interfaces and analytics while retaining limited commit authority. The NCI framework is complementary to these initiatives. It does not propose another maturity ladder. Instead, it provides a technology-agnostic characterization space for comparing how standardized components, industrial systems, and research prototypes differ in their concrete control-intelligence properties.

Several prior surveys review important parts of this landscape, including software-defined networking and network softwarization~\cite{kreutz2014software}, network function virtualization~\cite{zhang2020nfv}, machine learning for networking and cognitive network management~\cite{boutaba2018comprehensive,ayoubi2018machine}, network digital twins~\cite{almasan2022network,wu2021digital}, and recent applications of LLMs to networking~\cite{huang2024large}. These works provide broad coverage of enabling technologies and application domains. This paper takes a different perspective. It uses NCI to compare how successive generations of network-control systems strengthened decision-making, adaptation, operational knowledge, authority allocation, and operator interaction, and where their outputs still remained advisory, scoped, or constrained before changing network state. This lens structures the historical analysis that follows, tracing the uneven accumulation of control-intelligence properties across the three eras.


\begin{figure*}[!tb]
\begin{center}
\includegraphics[width=0.9\textwidth]{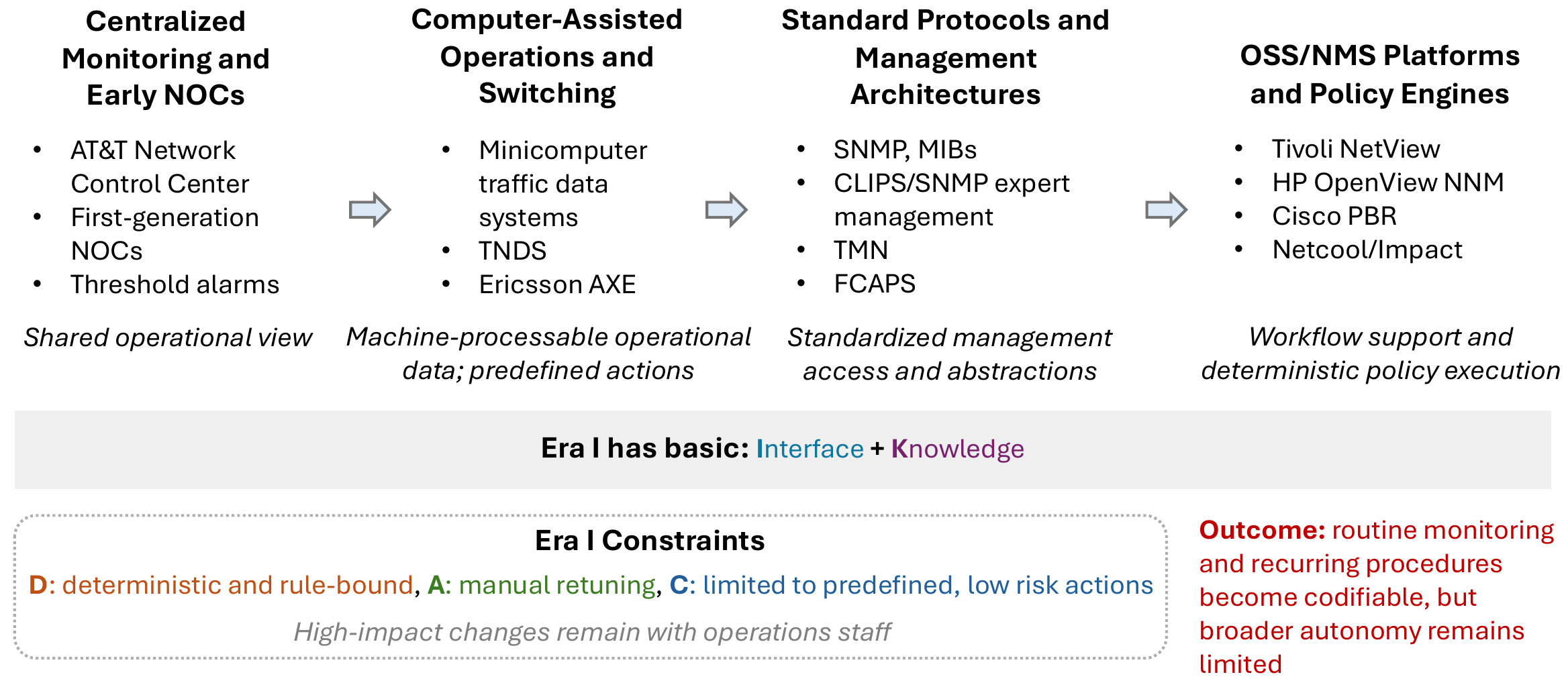}
\caption{Era-I progression from centralized monitoring to rule-based operational execution. The dominant advances were along the Interface and Knowledge axes: network state became more visible, machine-processable, standardized, and actionable through operations platforms. Control Delegation expanded only for predefined actions, while Decision Logic remained deterministic, and Adaptability depended on manual retuning.}
\label{fig:era1-progression}
\end{center}
\end{figure*}

\section{Rule-Based, Scripted Automation}
\label{sec:era-I}

The 1960s--early 2000s formed the operational foundation for modern network control. In this era, network operations shifted from largely manual, human-centered workflows to Network Operations Centers (NOCs), centralized monitoring, computer-assisted data collection, standardized management protocols, and deterministic, rule-based automation. Computerized systems enabled more observable network state and more consistent execution of predefined operational procedures. However, complex tasks and high-stakes decisions, such as incident interpretation, remedy selection, and approval of high-impact changes, were still managed by operators.

Before the 1970s, large-scale telecommunications operations relied heavily on labor-intensive, paper-based workflows, including manual record-keeping, shift reports, physical inspections, and reactive maintenance~\cite{bell1983engineering,dawson2000operations}. Although electromechanical control and alarm mechanisms already existed, they were typically attached to specific equipment, circuits, or fault conditions without a network-wide control process. The network's operative ``intelligence'' still relied largely on human expertise with little autonomous interpretation or decision-making. The transition toward rule-based network management unfolded gradually as centralized monitoring, traffic-data systems, software-controlled switching, standard management protocols, and operations platforms entered operational practice. This section examines this first era as the operational foundation of network automation, including centralized monitoring, traffic data systems, software-controlled switching, management protocols, and early OSS/NMS platforms, which made network operations more observable and repeatable. The cumulative progression of this era is summarized in Fig.~\ref{fig:era1-progression}.

\subsection{From Local Oversight to Centralized Monitoring}
AT\&T's opening of a Network Control Center in New York in 1962 was an early milestone in centralized network monitoring. The center addressed the growing complexity of long-distance telephone operations by moving beyond local oversight of geographically dispersed network elements. It provided centralized, near-real-time visibility into switching, routing, and traffic-status information for major toll network elements~\cite{att_history_mgmt}. This centralized view anticipated the operating pattern of later Network Operations Centers (NOCs), where operations staff could monitor distributed infrastructure from a shared operational picture.

\begin{table*}[t]
\centering
\footnotesize
\caption{NCI profile of representative Era-I developments. Scores are approximate qualitative positions within the five-axis framework: D = Decision Logic, A = Adaptability, K = Knowledge, C = Control Delegation, and I = Interface.}
\label{tab:era1-fiveaxis}
\setlength{\tabcolsep}{3.5pt}
\renewcommand{\arraystretch}{1.15}
\begin{tabularx}{\textwidth}{@{}p{0.17\textwidth}p{0.25\textwidth}cccccX@{}}
\toprule
\textbf{Development} & \textbf{Representative mechanisms} & \textbf{D} & \textbf{A} & \textbf{K} & \textbf{C} & \textbf{I} & \textbf{NCI synopsis} \\
\midrule
Centralized monitoring
& AT\&T Network Control Center; first-generation NOCs; threshold alarms
& D0--D1 & A0 & K1 & C0--C1 & I1
& Local inspection moved toward shared operational visibility; knowledge remained descriptive and alarm-centered. \\

Computer-assisted control
& Minicomputer-based traffic-data systems; TNDS; Ericsson AXE
& D0--D1 & A0 & K1--K2 & C1 & I1
& Operational data became more machine-processable, and network control actions became executable under predefined logic. \\

Standard protocols and management architectures
& SNMP; MIBs; CLIPS/SNMP expert management; TMN; FCAPS
& D1 & A0 & K2 & C0--C1 & I1--I2
& Managed objects, alerts, rule state, and management functions became more explicit and portable across systems. \\

Operations platforms and policy engines
& IBM Tivoli NetView; HP OpenView NNM; Cisco PBR; Netcool/Impact
& D1 & A0 & K2 & C1 & I2
& Standardized management information became actionable through workflow support and deterministic policy execution. \\
\bottomrule
\end{tabularx}
\end{table*}

First-generation NOCs used large electronic status boards and displays to present network status, allowing operations staff to track conditions through visual indicators and decide when to intervene. As software-based monitoring matured, threshold-based alarms and predefined event-handling rules reduced the manual monitoring burden. Specific events or metric thresholds could trigger predefined notifications or low-risk actions~\cite{boucadair2020framework}. Compared to paper-based and equipment-local practices, centralized monitoring advanced the \emph{Interface} axis from local inspection toward a shared operational view, and the \emph{Knowledge} axis from dispersed equipment signals toward consolidated network-state evidence. It also contributed to \emph{Control Delegation} by enabling the automatic triggering of alarms and low-risk notifications, though non-routine diagnosis and intervention remained with operations staff.

\subsection{From Manual Procedures to Computer-Assisted Control}
In the early 1970s, minicomputers began moving network operations from manual procedures toward computer-assisted control. By the mid-1970s, minicomputer-based systems could collect and process operational data in support of traffic analysis, engineering, and management. Bell System's Total Network Data System (TNDS), whose planning began in the mid-1960s and culminated in a family of systems documented in 1983~\cite{ebner1983total}, mechanized portions of traffic-data collection and processing. Digitally controlled switches also began to replace traditional electromechanical systems in the telephone infrastructure. Ericsson's AXE system, a telephone exchange deployed in the late 1970s, used modular software control to manage calls and network resources more flexibly~\cite{axe}. 

Compared with centralized monitoring, these systems made operational knowledge more machine-processable. TNDS mechanized traffic-data collection and processing, while AXE moved selected switching functions into software-controlled execution. From the NCI perspective, they advanced the \emph{Interface} axis from visual operational displays toward computer-assisted analysis, and the \emph{Knowledge} axis from centrally displayed state toward operational data that could be stored, processed, and reused. \emph{Control Delegation} also increased slightly, but only for predefined operational or switching actions.

\subsection{Standard Protocols and Management Architectures}
Another major development in this era was the standardization of network management protocols, with the Simple Network Management Protocol (SNMP) becoming the de facto standard for querying network devices and, where permitted, setting writable management variables in the late 1980s and early 1990s~\cite{case1988simple,case1990rfc1157}. Emerging from the Internet community's effort to unify management across a rapidly expanding infrastructure, SNMP evolved from the Simple Gateway Monitoring Protocol into a general manager/agent model for monitoring and limited control. The early SNMP specifications (RFC 1065-1067 and later RFC 1155-1157) defined the manager/agent model and basic protocol operations, including reading and setting device parameters and receiving asynchronous TRAP alerts~\cite{case1988simple,case1990rfc1157}. Together with Management Information Bases, SNMP enabled reactive, vendor-agnostic monitoring and broadened the reach of centralized, software-driven network management~\cite{perkins1997understanding}.

Early operational expert systems in IP networks coupled SNMP telemetry with CLIPS (a forward-chaining rule engine) to codify operational knowledge as production rules. NASA's implementation polled SNMP-managed elements for status and performance metrics, stored the returned data for rule evaluation, and applied problem-detection and problem-correction rules to adjust selected parameters or produce diagnostic output~\cite{faul1991using}. Architecturally, the CLIPS working memory and data files served as an inspectable state shared between SNMP access routines and the rule base, enabling repeatable yet rule-bound operations.
The need for IP network management also catalyzed lightweight monitoring tools: Multi Router Traffic Grapher used SNMP data to chart traffic statistics~\cite{oetiker1998mrtg}, while Big Brother checked distributed systems and emitted operational alerts for monitored services~\cite{sittler1997linux}.

The formalization of network management frameworks constituted another critical foundation for control. ITU-T Recommendation M.3010 defined the Telecommunications Management Network (TMN) reference architecture and its layered management model, spanning element, network, service, and business management abstractions~\cite{itu-t-m.3010}. ITU-T Recommendation M.3400 then specified TMN management functions across performance, fault, configuration, accounting, and security management, aligning closely with the fault, configuration, accounting, performance, and security (FCAPS) categories used in telecom-management practice~\cite{itu-t-m.3400}.
With respect to earlier computer-assisted systems, SNMP, MIBs, CLIPS-based rules, and TMN/FCAPS frameworks made the \emph{Interface} and \emph{Knowledge} axes less device-local and less vendor-specific. Managed objects, telemetry access, alerts, rule state, and management functions could be represented through more explicit protocols and architectural structures. This also supplied the standardized management layer on which later OSS/NMS platforms and policy engines could operate.

\subsection{Operations Platforms and Policy Engines}

Early rule-based automation matured along two converging tracks. OSS and NMS platforms consolidated alarms, topology, and workflows across FCAPS functions, while policy engines encoded deterministic, pre-approved behaviors in management, control, or forwarding processes. Together, these tracks moved network operations from isolated monitoring and manual procedures toward rule-governed operational workflows.
OSS/NMS platforms such as IBM Tivoli NetView~\cite{tivoli} and Hewlett-Packard OpenView Network Node Manager (NNM)~\cite{hp_nnm_manual,hp_nnm_scale} exemplified this shift. They supported centralized alarm handling, topology-aware monitoring, event forwarding, and operator workflow support, often building on standard management protocols and architectures such as SNMP and TMN. They consolidated operational context into platforms through which operations staff could inspect events, correlate symptoms, and coordinate responses.

Meanwhile, deterministic policy mechanisms began to move selected decisions closer to managed devices and event streams. Policy engines in this era appeared both inside devices and at the OSS/NMS layer. With Cisco IOS Release 11.0 in the mid-1990s, Policy-Based Routing (PBR) was operationalized as an on-device policy interpreter that used route maps to match packet attributes, such as IP access-list membership or packet length, and then set a next hop or output interface. This allowed routers to send selected packets along paths other than the default shortest path~\cite{cisco_pbr}. Netcool/Impact 3.1, documented in the mid-2000s, connected event sources to a rule-driven policy engine for event handling and contextual enrichment. Upon receiving an alarm, ``Impact" could apply filters, run policies, enrich events with contextual data, correlate or suppress alerts, and invoke configured automations through management integrations~\cite{ibm_impact}.

These systems expanded the scale and consistency of rule-based operations. They made alarms, topology, and event context more accessible through management platforms, and they allowed predefined rules to trigger selected forwarding or event-handling actions. Compared with earlier centralized monitoring, they advanced the \emph{Interface} axis from display-oriented visibility toward workflow support, and the \emph{Knowledge} axis from isolated device status toward consolidated alarms, topology, and event context.

\subsection*{NCI Synthesis of this era}

This era established the operational foundation of modern network control. Centralized monitoring and NOCs made distributed network state visible from a shared operational view; traffic-data systems and software-controlled switching made selected operational data and actions machine-processable; SNMP, MIBs, TMN, and FCAPS categories standardized management interfaces and operational abstractions; and early OSS and NMS platforms connected alarms, topology, workflows, and deterministic policies. These developments made routine monitoring feasible and allowed routine procedures to be codified as scripts, runbooks, and event-handling rules.

From the five-axis perspective, the systems in this era occupy a low-autonomy, low-adaptation region of the NCI space, as summarized in Table~\ref{tab:era1-fiveaxis}. \textbf{Interface} and \textbf{Knowledge} advanced most visibly. Interfaces evolved from local inspection and status displays toward centralized dashboards, management protocols, OSS/NMS consoles, and workflow support. Knowledge evolved from dispersed equipment signals and paper records toward machine-readable counters, alarms, topology information, managed objects, and management-function categories. However, this knowledge remained primarily descriptive rather than predictive or causal, and was usually organized around devices, links, alarms, and procedures instead of service-level intent or cross-domain models.

The remaining axes advanced more modestly. \textbf{Decision Logic} was predominantly reactive and deterministic, relying on threshold triggers, route maps, event filters, production rules, and scripted procedures that produced local or one-step responses rather than explicit planning or multi-objective reasoning. \textbf{Adaptability} was limited because changes in traffic patterns, failures, or operational policies usually required manual retuning of thresholds, scripts, rules, and workflows. \textbf{Control Delegation} remained tightly scoped: automated mechanisms could trigger routine notifications, execute predefined event-handling actions, or apply selected device-level policies, but non-routine incidents and high-impact changes still required operations-staff judgment, review, and approval.

The limiting factor of this era was not the absence of automation. Rather, the automation mechanisms lacked expressive decision logic, safe adaptation under non-stationary conditions, rich operational state models, and assurance mechanisms to circumscribe the risk of autonomous control actions. These limitations motivated the next era of network control, as network scale, heterogeneity, and service dynamism exceeded what centralized monitoring, deterministic rules, static workflows, and manual retuning could sustain.


\section{Programmable and Data-Driven Control}
\label{sec:era-II}

\begin{figure*}
\centering
\includegraphics[width=1\textwidth]{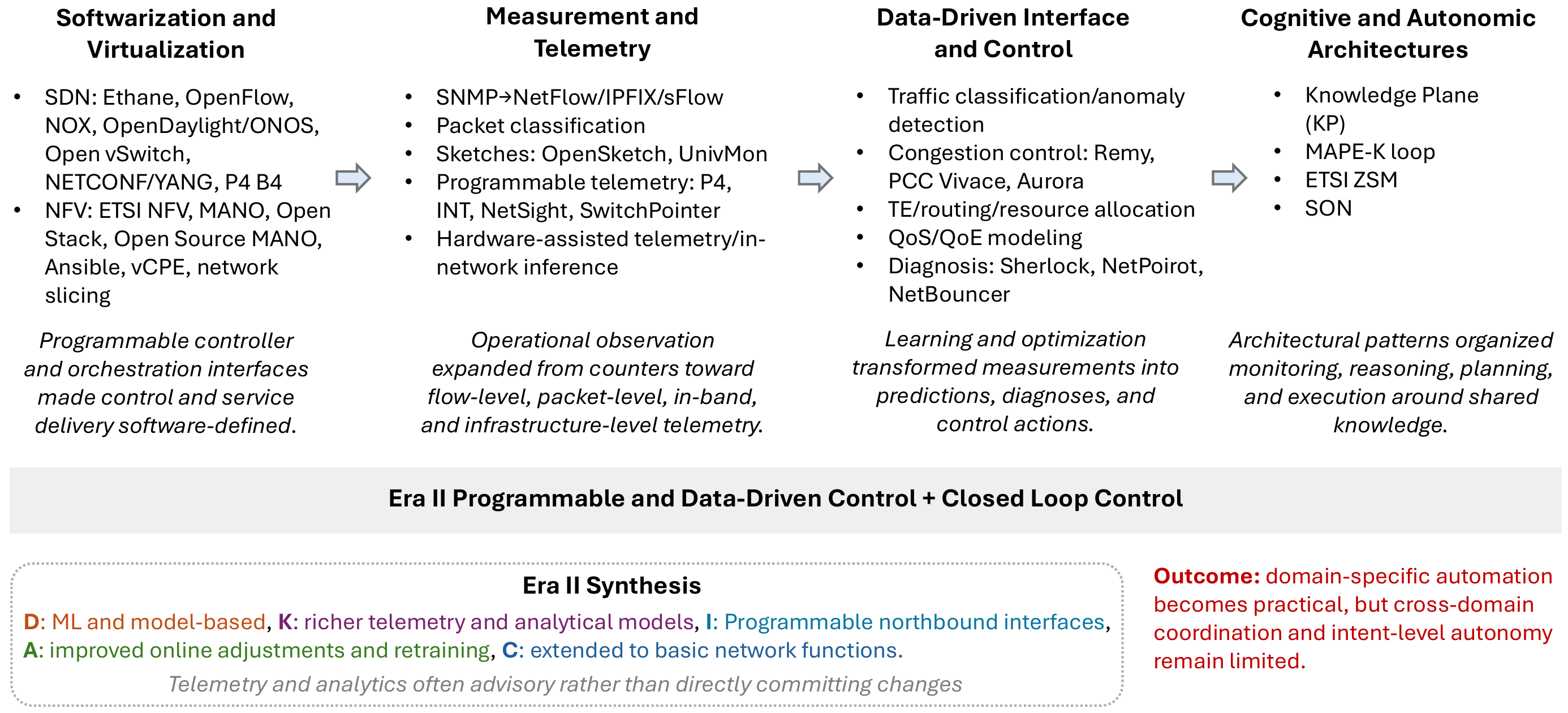}
\caption{Era-II progression from programmable actuation to data-driven closed-loop control. Network softwarization and virtualization exposed controller and orchestration interfaces; measurement and telemetry expanded operational knowledge from counters toward flow-level, packet-level, in-band, and infrastructure-level observations; data-driven inference transformed measurements into predictions, diagnoses, and optimization inputs; and cognitive/autonomic architectures organized these capabilities into closed-loop control patterns. Through the NCI lens, the era produced strong advances in \emph{Decision Logic}, \emph{Knowledge}, and \emph{Interface}, with selected gains in \emph{Adaptability} and bounded expansion of \emph{Control Delegation}. The residual need to coordinate domain-specific controllers, telemetry pipelines, and learned models sets the stage for the LLM-enabled era.}
\label{fig:era2-progression}
\end{figure*}

From the 2000s to the mid 2020s, network control evolved from device-local scripts and rule engines toward programmable and data-driven control loops supported by increasingly fine-grained and timely telemetry. Compared with the rule-based and scripted era, the control and service-delivery layer became more programmable. Forwarding behavior could be controlled through logically centralized controllers. Network functions could be instantiated and orchestrated as software in parallel, with operational observation expanding from periodically polled device counters to flow-level, packet-level, in-band, and infrastructure-level measurements, while learning-based analytics increasingly supported prediction, diagnosis, and optimization. This section traces these evolutions through network softwarization and virtualization, measurement and telemetry, data-driven analytics, and cognitive and autonomic architectures that organize monitoring, reasoning, and actuation into closed-loop control, as depicted in Fig.~\ref{fig:era2-progression}.


\subsection{Programmable Actuation through Network Softwarization}
Since the mid 2000s, network softwarization and virtualization have reshaped how control logic and network functions are implemented. In particular, Software-Defined Networking (SDN) separated control logic from packet forwarding and exposed programmable controller interfaces, while Network Functions Virtualization (NFV) decoupled network functions from dedicated appliances and introduced software-based lifecycle management over shared infrastructure.

\subsubsection*{Software-Defined Networking}
SDN introduced the decoupling of the control plane (decision logic) from the data plane (packet-forwarding fabric), thereby enabling logically centralized control~\cite{kreutz2014software}. The intellectual roots of SDN date back to the ``active networks" initiative in the late 1990s~\cite{tennenhouse1996towards}, which proposed in-network programmability through code injection. That initiative was later discontinued because of practical deployment constraints~\cite{feamster2014road}. The modern SDN movement began with the Ethane project at Stanford in 2007, which demonstrated the benefits of separating control logic from switch hardware in enterprise networks~\cite{casado2007ethane}. This project led to the OpenFlow protocol~\cite{mckeown2008openflow}, which defined the standard interface for controllers to program forwarding devices. These developments reduced dependence on vendor-specific control planes and catalyzed a broader movement toward open programmability.
Once that architectural split was established, the surrounding ecosystem supplied controllers, emulators, virtualization layers, and programming languages.
Early SDN systems permeated the operational control stack: NOX supported global coordination~\cite{gude2008nox}, FlowVisor enabled isolation-preserving slicing~\cite{sherwood2009flowvisor}, Mininet made large-scale emulation accessible~\cite{lantz2010network}, and Frenetic introduced higher-level programming abstractions~\cite{foster2011frenetic}. VMware's acquisition of Nicira in 2012~\cite{nicira} signaled the commercial potential of SDN, and Google's B4 deployment~\cite{jain2013b4} showed that centralized traffic engineering could operate at scale in a private WAN.

These advances greatly galvanized institutional and operational consolidation. In 2011, the Open Networking Foundation (ONF) was launched by major network operators and web-scale companies, with vendors joining as additional members to promote SDN standardization and interoperability~\cite{parulkar2011open}. Projects such as OpenDaylight~\cite{odl}, ONOS~\cite{onos}, and Open vSwitch~\cite{pfaff2015design} broadened the SDN software ecosystem across controllers and programmable switching. In contrast, the open disaggregation movement and systems such as SONiC~\cite{sonic2017} made vendor-neutral programmability a practical design objective.
During the 2010s, interfaces such as NETCONF, standardized earlier and widely adopted alongside YANG-based management, complemented OpenFlow with persistent configuration management~\cite{enns2011network,claise2019network}, while P4 extended programmability directly into the data plane~\cite{bosshart2014p4}.

\subsubsection*{Network Function Virtualization}

NFV transformed service provisioning through virtualization and cloud technologies. Its origins are commonly traced to a seminal ETSI position paper published in 2012, in which a consortium of global CSPs formally proposed replacing traditional middleboxes with Virtual Network Functions (VNFs) running on commercial off-the-shelf (COTS) servers~\cite{isg2013network}.
ETSI then established the NFV Industry Specification Group (ISG) to expedite standardization, while early deployments such as virtual Customer Premises Equipment (vCPE) demonstrated operational feasibility~\cite{cpe}. Telcos adopted cloud platforms such as OpenStack as Virtual Infrastructure Managers (VIMs)~\cite{openstack}, and the NFV 001--004 specifications introduced the Management and Orchestration (MANO) framework for lifecycle control, orchestration, and OSS/BSS integration~\cite{etsi-nfv}. Open-source implementations such as Open Source MANO~\cite{openmano} and playbook-style automation such as Ansible~\cite{ansible-github} provided practical mechanisms for repeatable VNF and device lifecycle actions.

NFV also exposed a practical constraint for autonomous control: virtualized services can be orchestrated effectively only when their data-plane performance remains predictable. This motivated optimized packet-I/O and software data-plane techniques, including kernel-bypass frameworks, poll-mode drivers, batching, hugepages, high-performance virtual switching, and, later, in-kernel fast paths such as eBPF/XDP and AF\_XDP~\cite{rizzo2012netmap,zhang2020nfv,dpdk,pfaff2015design,hoiland2018xdp,parola2023comparing}. These techniques are not themselves NCI mechanisms, but they made virtualized network functions sufficiently performant to participate in programmable orchestration loops. By the second half of the 2010s, operator programs such as AT\&T's Domain 2.0~\cite{att2014domain2} and Telef{\'o}nica's UNICA platform~\cite{telefonica2016unica} had framed large-scale virtualization and orchestration as central telecom transformation objectives.
Notably, the programmability introduced by SDN/NFV also enabled network slicing, in which compute, transport, and network function resources are assembled and managed as logically isolated service instances. In 5G systems, this made the dynamic allocation and lifecycle management of virtualized resources a central objective of orchestration~\cite{foukas2017network}.


With SDN and NFV, control actions no longer need to be embedded only in device-local procedures; they can be explicitly programmed through software interfaces. Compared with the rule-governed management era, this advanced the \emph{Interface} axis from management consoles and protocols toward controller and orchestration APIs, and the \emph{Decision Logic} axis from local rules toward centralized policies, optimization programs, and lifecycle-management workflows.

\subsection{Richer Knowledge through Measurement and Telemetry}
While SDN and NFV established programmable actuation, a parallel evolution expanded what network-control systems could observe. Measurement and telemetry progressed from periodic counter polling toward packet-level, flow-level, in-band, and infrastructure-level instrumentation.

\subsubsection*{From counters to packet- and flow-level telemetry}
Traditional mechanisms such as SNMP exposed management variables and counters through a polling-oriented management model~\cite{case1990rfc1157}. Operators could infer end-to-end behavior from such counters, but sparse samples were unfit for transient microbursts or millisecond-scale congestion in multi-tenant cloud environments. In response, the early 2000s saw the rise of flow-export protocols that bridged the gap between per-device counters and per-packet traces. Cisco's NetFlow and IETF's IPFIX~\cite{claise2013specification} introduced structured flow records summarizing traffic by 5-tuples, while sFlow~\cite{phaal2001inmon} provided statistically sampled packet and counter data. These mechanisms enabled scalable traffic monitoring across carrier backbones and data centers. The shift from periodic polling to asynchronous flow export marked the first step toward continuous telemetry, reducing data collection latency from minutes to seconds and laying the foundation for more advanced analysis~\cite{zhang2019flowatcher}.

\subsubsection*{Packet classification and header processing}
As flow-level telemetry matured, operators sought finer visibility into sub-flow behaviors, such as application classes, policy domains, or security signatures. This need made \emph{packet classification} a core dataplane primitive, in which packets had to be matched against large, multidimensional rule sets at line rate. Early work moved beyond linear search and hashing through tuple-space indexing~\cite{srinivasan1999tuple}, hierarchical geometric partitioning~\cite{gupta2000hicuts,singh2003hypercuts}, and bit-parallel representations~\cite{baboescu2001scalable}.

As packet processing moved into high-performance software dataplanes, rule update efficiency became the primary concern. Classifiers had to sustain line-rate lookups while absorbing frequent rule changes driven by policy and service churn. TupleMerge~\cite{daly2019tuplemerge} revisited tuple-space methods for this online setting, and NPC~\cite{zhang2025npc} made classifier construction network-aware by incorporating measured traffic characteristics. Together, these methods show how classification co-evolved with measurement and control in programmable dataplanes, enabling selective mirroring, per-class telemetry, and event filtering at high packet processing rates.

\subsubsection*{In-network computing}
As programmable data planes matured, a set of measurement and classification functions can be recast as lightweight in-network inference under strict pipeline constraints. Iisy~\cite{zheng2024iisy}, Planter~\cite{zheng2021planter}, and Homunculus~\cite{swamy2023homunculus} demonstrated this trajectory by mapping classifiers or machine-learning pipelines onto programmable devices subject to constraints on features, state, latency, and throughput.

\subsubsection*{Sketch-based streaming measurement}
To capture statistical traffic characteristics without maintaining per-flow state, researchers introduced probabilistic sketches, compact data structures that approximate counts, frequencies, or cardinalities within fixed memory; HyperLogLog is a representative cardinality estimator~\cite{flajolet2007hyperloglog}. OpenSketch~\cite{yu2013opensketch} and UnivMon~\cite{liu2016one} integrated sketching with programmable data paths, while subsequent systems addressed constraints on configuration, implementation, and reconfiguration~\cite{huang2018sketchlearn,namkung2022sketchlib,zhou2022sketchguide}. These efforts bridged the gap between software and hardware telemetry, showing that streaming summarization could coexist with packet-level instrumentation.

\subsubsection*{Programmable data-plane telemetry}
In parallel with advances in sketch-based summarization, programmable switching enabled direct, in-band measurement within forwarding pipelines. P4 enabled programmable packet parsing and header processing across protocol-independent packet processors~\cite{bosshart2014p4}, and In-band Network Telemetry (INT) demonstrated how packets could carry hop-by-hop metadata such as timestamps, queue occupancy, and egress ports~\cite{kim2015intdemo}. Complementary postcard or report-based schemes, such as NetSight~\cite{handigol2014packethist} and SwitchPointer~\cite{tammana2018switchpointer}, collect per-hop visibility without requiring the same packet-header modification model as INT. Hybrid designs then combined in-band telemetry with sketches or sampling techniques~\cite{sivaraman2016packet, basat2020pint}. Together, these systems made fine-grained packet-level introspection practical for operational monitoring and feedback control.

\subsubsection*{Hardware-assisted telemetry}
Beyond packet-centric and in-band mechanisms, a complementary line of inquiry examined {hardware-assisted telemetry}, leveraging low-level device and microarchitectural signals to indicate network dynamics. Dobrescu~et al.~\cite{dobrescu2012toward} helped establish this direction by instrumenting software packet-processing platforms to expose system-level signals as predictors of performance. Their work demonstrated that observing fine-grained system events can support inference of throughput and latency variability without explicit packet sampling, thereby bridging the gap between host-level instrumentation and end-to-end telemetry. Building on this principle, later systems exploited host-level counters, queueing signals, and infrastructure-level performance indicators to infer contention and service degradation in virtualized network functions~\cite{dobrescu2012toward, tootoonchian2018resq, manousis2020contention}. Together, these works revealed a complementary trend in telemetry: they treated the underlying infrastructure itself as a source of operational observability, beyond exporting packet statistics alone. This systems-resident perspective links packet-level measurements to performance inference, providing low-overhead insights valuable for network automation.

In summary, flow export, packet classification, sketches, programmable telemetry, hardware-assisted measurement, and bounded in-network inference expanded the operational evidence available to network-control systems. Compared with the counters and alarms of the rule-based and scripted era, these mechanisms provided finer-grained, timelier, and more policy-relevant representations of network state. Their principal NCI contribution was along the \emph{Knowledge} axis, supplying the evidence needed for anomaly detection, traffic engineering, capacity planning, diagnosis, and feedback control. They improved what a control system could observe and infer.

\subsection{Data-Driven Inference and Learning-Based Control}
As richer telemetry became available since the 2010s, machine learning (ML) techniques, especially Deep Learning (DL) and Reinforcement Learning (RL), have been increasingly adopted to support inference, prediction, diagnosis, and optimization in network operations~\cite{boutaba2018comprehensive,ayoubi2018machine}. In application domains such as traffic classification, congestion control, traffic engineering, resource allocation, and fault diagnosis, ML models have transcended traditional fixed-threshold rules by estimating hidden states, forecasting future conditions, and, in some cases, recommending or selecting control actions.

Data-driven control models were first widely applied in traffic classification and cybersecurity analytics~\cite{moore2005internet,pacheco2018towards,lakhina2004characterization}. Traditional approaches based on static port numbers or signature-based deep packet inspection became less reliable as encryption and application multiplexing increased, creating demand for classifiers that could infer application or traffic categories from statistical flow features~\cite{anderson2017machine}.
ML-based classifiers learned traffic classes from flow features~\cite{moore2005internet}, while later neural models learned packet representations with less manual feature engineering~\cite{lotfollahi2020deep} and generated compact classification structures for specific rule sets~\cite{liang2019neural}. Related supervised and unsupervised methods supported anomaly and intrusion detection~\cite{mirsky2018kitsune}. Through the NCI lens, these systems advanced the \emph{Knowledge} and \emph{Decision Logic} axes by transforming raw traffic measurements into inferred application classes, traffic types, and security-relevant labels.

Beyond traffic classification and security analytics, data-driven control models also reshaped congestion control at the transport layer. Classical Transmission Control Protocol (TCP) variants such as Reno~\cite{jacobson1988congestion} and CUBIC~\cite{ha2008cubic} relied on handcrafted control rules. Remy~\cite{winstein2013remy} showed that congestion-control rules could be synthesized through offline optimization; PCC Vivace~\cite{dong2018pccvivace} combined online learning with convex optimization to adapt sending rates from observed loss and latency; and Aurora~\cite{jay2019aurora} used deep reinforcement learning to adjust congestion windows from network feedback. These systems marked a shift from fixed heuristics to controllers that adapt sender behavior based on measured performance. Through the NCI lens, they advanced \emph{Decision Logic} by replacing fixed heuristics with learning-based optimization, and advanced \emph{Adaptability} by allowing sender behavior to change in response to measured loss, latency, and throughput. 

\begin{table*}[t]
\centering
\footnotesize
\caption{NCI profile of programmable and data-driven control developments. Scores indicate approximate qualitative positions within the five-axis framework: D = Decision Logic, A = Adaptability, K = Knowledge, C = Control Delegation, and I = Interface. Ranges reflect variation among the representative mechanisms grouped in each row.}
\label{tab:era2-fiveaxis}
\setlength{\tabcolsep}{3.5pt}
\renewcommand{\arraystretch}{1.15}
\begin{tabularx}{\textwidth}{@{}p{0.18\textwidth}p{0.25\textwidth}cccccX@{}}
\toprule
\textbf{Development} & \textbf{Representative mechanisms} & \textbf{D} & \textbf{A} & \textbf{K} & \textbf{C} & \textbf{I} & \textbf{Main NCI reading} \\
\midrule

Softwarization and virtualization
& SDN/OpenFlow; controllers; NFV MANO; VNF lifecycle orchestration
& D1 & A0--A1 & K1--K2 & C1--C2 & I1
& Advanced \emph{Interface} and \emph{Decision Logic} by exposing programmable controller and orchestration functions, while enabling scoped execution within controller-defined domains. \\

Measurement and telemetry
& NetFlow/IPFIX; sFlow; packet classification; sketches; INT; hardware-assisted telemetry
& D0 & A0 & K0--K1 & C0 & I1
& Advanced \emph{Knowledge} by making traffic, forwarding, queue, and infrastructure state more fine-grained and timely; the mechanisms remained primarily observational. \\

Data-driven analytics and control
& ML-based traffic classification; congestion control; TE; resource allocation; QoE; diagnosis; learned surrogates
& D1--D2 & A1--A2 & K1--K2 & C0--C2 & I1
& Advanced \emph{Knowledge} and \emph{Decision Logic} by converting measured state into inferred, predictive, and optimization-relevant representations; \emph{Adaptability} is improved in selected control loops. \\

Cognitive and autonomic architectures
& Knowledge Plane; MAPE-K; SON; ZSM; NWDAF analytics
& D1 & A1 & K2 & C0--C2 & I1--I2
& Made the composition of \emph{Knowledge}, \emph{Decision Logic}, and \emph{Interface} explicit; authority ranged from analytics-only services to bounded closed-loop actuation. \\

\bottomrule
\end{tabularx}
\end{table*}

Learning-based control has also expanded beyond sender-side adaptation to encompass network-wide traffic engineering (TE). In this setting, ML models were used to predict path performance and to support routing, scheduling, and resource allocation decisions over explicit network topologies and traffic demands. Specifically, ``Learning to Route"~\cite{valadarsky2017learning} reframed routing configuration as a learning problem, and subsequent work has combined graph neural networks (GNNs), reinforcement learning, and optimization. RouteNet~\cite{rusek2020routenet}, AuTO~\cite{chen2018auto}, Bernardez et al.~\cite{bernardez2021machine}, Teal~\cite{teal}, RedTE~\cite{gui2024redte}, and LO-TE~\cite{liu2025shooting} illustrate the progression from topology-aware prediction toward learned TE controllers. Through the NCI lens, these systems advanced \emph{Knowledge} by learning topology- and traffic-dependent performance models. They also advanced \emph{Decision Logic} by incorporating prediction, policy search, and optimization into routing decisions.

In addition to TE, data-driven control also supported resource allocation under stochastic demand. In cluster and data-center settings, Decima~\cite{mao2019learning} used graph-structured reinforcement learning for job-DAG scheduling. In wireless and 5G systems, GNN-based approaches modeled interference and resource dependencies: Shen et al.~\cite{shen2021gnnrrm} proposed a GNN architecture for scalable radio-resource management, while Sala{\"u}n et al.~\cite{salaun2022gnn} applied a GNN to cell-free massive-MIMO resource allocation. Through the NCI lens, these systems advanced \emph{Knowledge} by representing workload or interference dependencies as graphs, and advanced \emph{Decision Logic} and selected forms of \emph{Adaptability} by learning allocation policies that responded to changing workloads, topologies, and conditions.

Data-driven control also broadened the objectives represented in network decision loops. Quality-of-service (QoS) and quality-of-experience (QoE) management used machine learning to relate network-level measurements to user-perceived outcomes. Measurement studies showed that video quality could be inferred even from encrypted traffic~\cite{dimopoulos2016measuring}, while Pensieve~\cite{mao2017neural} used reinforcement learning to optimize adaptive-bitrate decisions under time-varying network conditions. Through the NCI lens, these systems advanced \emph{Knowledge} by estimating user-perceived service quality from network observations and advanced \emph{Decision Logic} by incorporating those estimates into control and action selection.

A parallel line of work used ML methods to diagnose abnormal behavior. In particular, Sherlock~\cite{bahl2007sherlock} modeled causal dependencies between faults and symptoms, NetPoirot~\cite{arzani2016netpoirot} distinguished network- from host-level faults in hyperscale data centers, and NetBouncer~\cite{tan2019netbouncer} combined active probing with latent-factor learning to localize device and link failures in production data-center networks. Learned surrogate models were also used to approximate expensive simulation and performance-analysis pipelines. For instance, MimicNet~\cite{zhang2021mimicnet}, DeepQueueNet~\cite{yang2022deepqueuenet}, and m3~\cite{li2024m3} learned surrogates for packet-level simulation, queueing behavior, or flow-level performance estimation. Through the NCI lens, these systems advanced the \emph{Knowledge} axis from measured and inferred network state toward explanatory and predictive domain models that could support fault localization, scenario evaluation, planning, and optimization.

Across these applications, data-driven models changed how operational evidence was interpreted and incorporated into network decisions. Compared with telemetry mechanisms that primarily exposed the measured state, these models inferred traffic classes, anomalies, likely root causes, performance, and candidate actions. The main NCI advances, therefore, occurred in \emph{Knowledge} and \emph{Decision Logic}: operational state became more predictive and explanatory, while decision-making increasingly incorporated learned models, policy search, and optimization. \emph{Adaptability} also improved in selected domains through online rate adjustment, periodic retraining, re-optimization, and controlled policy updates. These gains nevertheless remained domain-bounded: many systems provided decision support, while direct controllers operated within predefined objectives, state representations, and execution scopes rather than exercising broad operational authority.

\subsection{Closed-Loop Cognitive and Autonomic Architectures}

The preceding developments advanced specific capabilities, including programmable control, fine-grained observation, prediction, diagnosis, and optimization. Cognitive and autonomic architectures addressed a complementary question: how should these capabilities be organized around shared operational knowledge and connected to decision-making and actuation through observed feedback?

The Knowledge Plane (KP) offered an influential architectural vision~\cite{clark2003knowledge}. It envisaged a logically integrated yet physically distributed layer to maintain network-wide models, combine operational context, and inform the control/data planes. It frames network control around a knowledge layer to support inference, planning, and coordinated decision-making.

Autonomic computing provided a control pattern through the Monitor-Analyze-Plan-Execute-over-shared-Knowledge (MAPE-K) loop~\cite{kephart2003vision}. MAPE-K structured self-management around monitoring, analysis, planning, and execution functions connected through a shared knowledge base. Its agenda encompassed self-configuration, self-optimization, self-healing, and self-protection under high-level policies rather than continuous low-level intervention. Early autonomic-networking prototypes explored how distributed autonomic elements could share operational knowledge, invoke rule engines and optimizers, and execute actions~\cite{strassner2009icumt}. 

In essence, KP emphasized the network-wide organization of operational knowledge and reasoning, while MAPE-K specified a reusable feedback structure connecting observation, analysis, planning, and action. These ideas later influenced telecom-management architectures. Notably, ETSI Zero-touch Network and Service Management (ZSM) defined management services, closed-loop functions, and interfaces for coordinating automation across management domains~\cite{etsi-gs-zsm}. 
These principles also appeared in more domain-specific settings. Self-Organizing Network (SON) functions implemented local self-configuration and self-optimization loops~\cite{feng2008self}, whereas the Network Data Analytics Function (NWDAF) in 5G exposed analytics services to other network functions without directly committing network changes~\cite{3gpp.29.520}. 

Through the NCI lens, this line of work made the composition of control intelligence explicit. KP emphasized the integration of operational knowledge and reasoning; MAPE-K organized monitoring, analysis, planning, and execution around shared knowledge; ZSM translated related principles into a telecom management architecture with defined functional and interface boundaries; and SON and NWDAF illustrated how closed-loop control and analytics services could be instantiated within bounded mobile network domains. Collectively, these developments advanced \emph{Knowledge}, \emph{Decision Logic}, and \emph{Interface} by clarifying how observation, reasoning, and actuation could be connected. They also showed that \emph{Control Delegation} depends on where execution authority is allocated: SON functions may operate within restricted control scopes, whereas analytics functions, such as NWDAF, are primarily tasked to inform other network functions.

\subsection*{NCI Synthesis of this era}

In this era, the scope and capability of network control have been materially expanded through the integration of programmable actuation with richer operational state and data-driven decision support. Network softwarization and virtualization exposed controller and orchestration interfaces through which policies, optimization programs, and lifecycle-management workflows could affect network state. Measurement and telemetry supplied finer-grained and timelier representations of traffic, performance, faults, and resource usage. Data-driven models transformed these observations into inferred states, predictions, diagnoses, and optimization guidance. Cognitive and autonomic architectures then provided reusable patterns for composing observation, shared knowledge, analysis, planning, and execution into a closed-loop control system. Table~\ref{tab:era2-fiveaxis} summarizes this progression in the five-axis NCI space. 
The strongest advances occurred along \emph{Decision Logic}, \emph{Knowledge}, and \emph{Interface}. \emph{Decision Logic} moved beyond threshold rules and static scripts toward centralized policy programs, optimization-based controllers, learned models, and hybrid combinations of prediction and optimization. \emph{Knowledge} expanded from counters, alarms, and basic topology records toward streaming telemetry, inferred operational state, predictive domain models, and shared knowledge structures. \emph{Interface} evolved from device-oriented commands and management protocols toward programmable controller APIs, orchestration frameworks, telemetry services, and higher-level management abstractions.

Progress along \emph{Adaptability} and \emph{Control Delegation} was more uneven. Adaptability improved in systems that updated sending rates, routing decisions, resource-allocation policies, or learned models in response to changing conditions. Such adaptation was often confined to a particular control loop or implemented through periodic re-optimization, offline retraining, or controlled policy updates. Control Delegation also expanded into congestion control, traffic engineering, virtualized-service lifecycle management, SON functions, and closed-loop management frameworks. Notably, execution authority remained regulated by explicit objectives, policy scopes, controller-defined guardrails, and, for higher-impact actions, human supervision. Telemetry and analytics functions often remained advisory because they supplied operational evidence or recommendations without directly committing changes.

These advances led to substantial capability within individual control loops, but limited coordination across them. Programmable controllers, telemetry pipelines, and learned models were typically designed for particular domains, timescales, and operational tasks. Operators continued to translate high-level objectives into domain-specific inputs, reconcile evidence across heterogeneous systems, and coordinate multi-step workflows spanning multiple control and management domains. These residual interpretation and coordination burdens created a natural role for LLM-enabled network operations. LLMs can assist with intent interpretation, evidence synthesis, and workflow assembly, but they do not by themselves make adaptation safe or broader autonomy defensible. As discussed in Section~\ref{sec:era-III}, authoritative actuation consequently remains anchored in policy-governed controllers, verification mechanisms, authorization, staged rollout, and rollback.


\section{LLM-Enabled Network Operations}
\label{sec:era-III}

Since the early 2020s, the burdens of interpretation and coordination have become more pronounced with the proliferation of 5G network slicing, cloud-native network functions, and RAN disaggregation. Services and OAM responsibilities no longer align neatly with individual devices or single domains. Instead, they increasingly span the core, wide-area network (WAN), data center, and RAN~\cite{barakabitze20205g}. 
As a result, operational incidents and network changes have increasingly become cross-domain concerns. A configuration error or runtime anomaly in one domain may propagate through service dependencies, while multi-vendor environments expose heterogeneous data models, interfaces, and control semantics. Cloud-native deployment practices also increase the frequency of software and configuration changes, while operators must continue to satisfy reliability, compliance, and service-level constraints~\cite{capegemini-report}. The operational challenge thus extends beyond optimizing individual control loops within a single domain to interpreting intent, assembling cross-domain operational context, and, most importantly, reliably and consistently executing changes through auditable validation, authorization, rollout, and recovery mechanisms~\cite{GB1524A}.
Together, these operational pressures motivate the emergence of a third era of NCI. Programmable, data-driven control remains necessary, while network operations increasingly require intent interpretation, evidence-grounded analysis, cross-domain workflow coordination, and governed change execution~\cite{etsi-gs-zsm,etsi-zsm-009,etsi-zsm-016}.

\begin{figure*}[!t]
\centering
\includegraphics[width=0.9\textwidth]{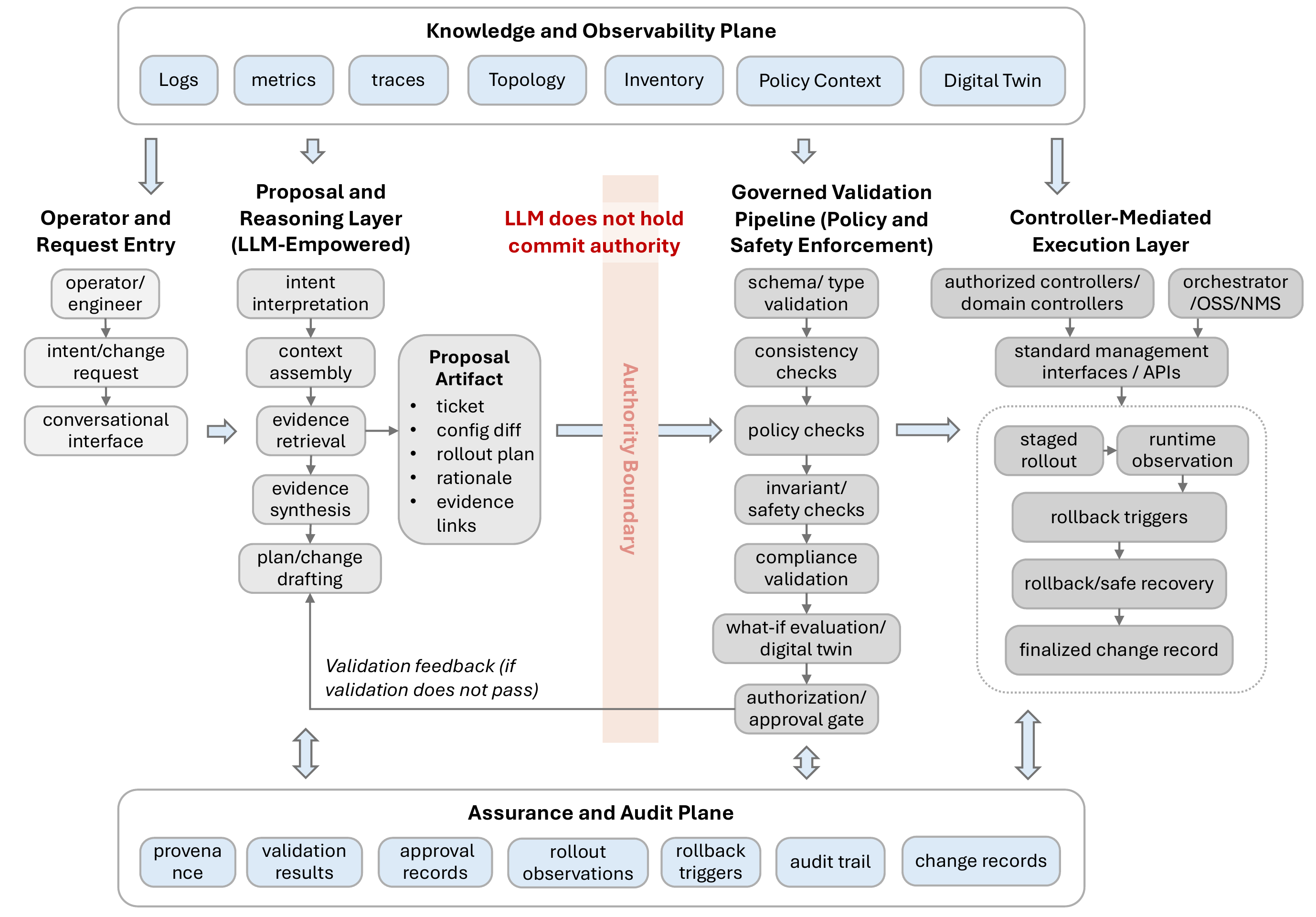}
\caption{Reference architecture for LLM-enabled governed network operations. The critical architectural boundary is between proposal generation and commit authority: the LLM assembles operational context and drafts a proposal artifact, while deterministic checks, authorization, staged rollout, and rollback remain in the governed execution path. Detailed pre-change checks can include schema/type validation, policy and invariant checks, reachability and compliance validation, and what-if evaluation in a simulator or digital twin.}
\label{fig:era3-refarch}
\end{figure*}

To tackle this operational backdrop, networking research and industry initiatives increasingly explored whether LLMs could reduce the burden of interpreting requests and coordinating OAM workflows, while continuing to rely on established procedures for validating, approving, and executing network changes~\cite{huang2024large,capegemini-report,deutsche}. Built on the transformer architecture~\cite{vaswani2017attention}, LLMs demonstrated exceptional capabilities in instruction following, code generation, text summarization, retrieval-based analysis, and multi-step tool use~\cite{brown2020language,chatgpt,team2023gemini}. These capabilities made them relevant to tasks that require interpreting intent, synthesizing heterogeneous evidence, and assembling workflows across existing control systems.

In practice, LLMs are most applicable to tasks performed before an operational change is approved and executed.
For example, an LLM-enabled system may interpret an operator request; retrieve and synthesize logs, tickets, runbooks, topology information, and configuration state; draft diagnostic hypotheses, proposed configuration changes, or rollout plans; and organize an OAM procedure. To support these tasks, the LLM is typically embedded in a workflow that supplies relevant context, exposes usable tools, preserves intermediate results, and records outputs for subsequent checking~\cite{yao2023react,huang2024large,jiang2025large,anthropic2025context,anthropic2025harnesses,openai2026harness}. When the workflow enables the LLM to pursue an objective by selecting tools, maintaining intermediate state, and planning subsequent steps, it is commonly described as an LLM-enabled agent.

To explain how these workflows interact with existing network control systems, this section distinguishes among three concepts. A \emph{proposal artifact} is an inspectable output prepared for review or validation, such as a diagnostic hypothesis, verification query, proposed configuration change, objective specification, or rollout plan. \emph{Commit authority} is the authority to approve a proposal for application to the operational network. \emph{Governed execution} denotes the controlled path from an approved proposal to an operational change. It includes validation and authorization before deployment, execution through established control or management systems, plus monitoring, rollback, and record-keeping afterward.
Under this distinction, LLM-enabled systems interpret operator requests, assemble relevant evidence from heterogeneous OAM records, translate intent into structured operational inputs, and produce proposals that existing validation and execution systems can process. Operators or established authorization processes determine whether a proposal may proceed, while existing control and management systems apply the approved change~\cite{wang2024netassistant,llmnetcfg}.

The separation between proposal generation and authorized execution also resonates in standards and industry guidance. In particular, TM Forum's introductory guide treats governance, oversight, traceability, interoperable interfaces, policy representation, and audit records as architectural concerns for AI-enabled closed loops~\cite{IG1251E, IG1414}. In cellular networks, 3GPP and O-RAN define management and control interfaces through which operational tools interact with established network functions~\cite{3gpp.29.520,o-ran}. Vendor and operator white papers similarly describe architectures in which higher-level functions interpret intent and coordinate service and resource management above domain controllers, supported by assurance functions, knowledge bases, digital twins, and human supervision~\cite{nokia-anf,ericsson-aws,huawei-agent,netcracker,deutsche,telefonica}. Likewise, Appledore argues that LLM-based components should be integrated into existing control loops for autonomous network operation rather than deployed as a separate automation system~\cite{appledore-agentic-control-loop}.

One related architectural consideration is the distinction between fast and slow control loops. Latency-critical control is generally better handled by deterministic controllers, compact learned policies, or optimization-based mechanisms whose timing and stability can be bounded. LLM-enabled components are more naturally placed in slower or supervisory workflows, where they can support intent interpretation, the synthesis of telemetry and logs, policy analysis, and the coordination of OAM procedures~\cite{anm}.

\begin{figure*}[!tb]
\centering
\includegraphics[width=0.9\textwidth]{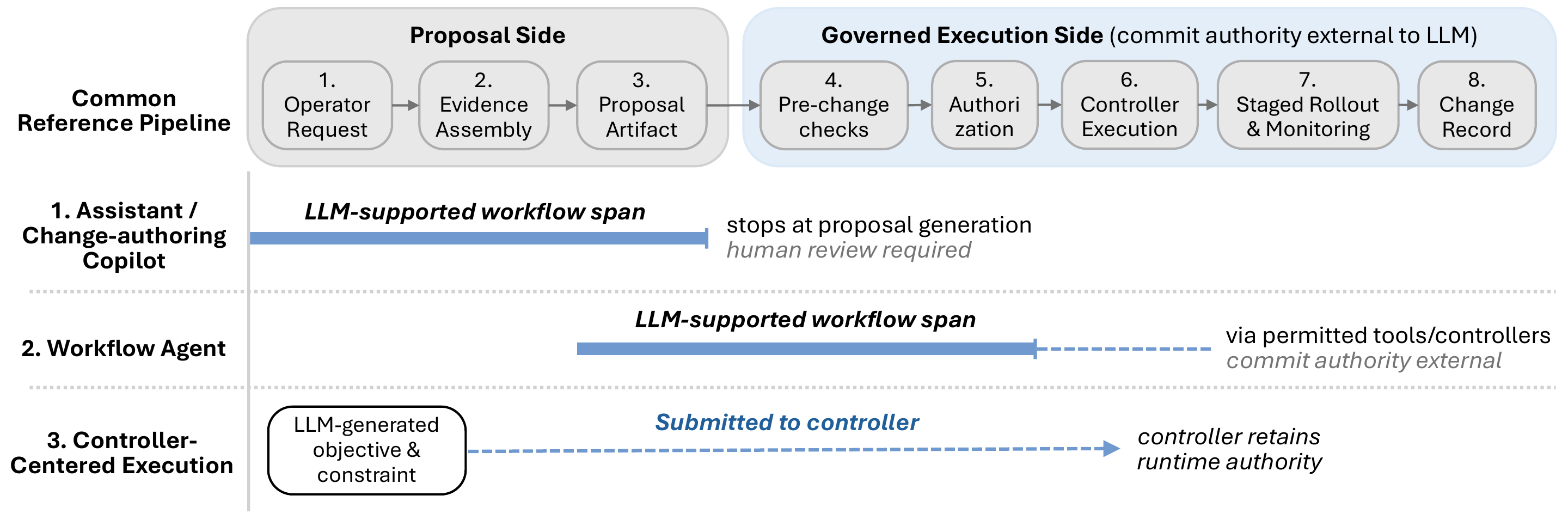}
\caption{Placement of the three integration patterns over the common operational pipeline. Their main architectural difference is not merely whether an LLM is present, but where LLM-supported processing terminates relative to the commit-authority boundary: assistants remain on the proposal side, OAM workflow agents reach permission-scoped procedure execution, and controller-centered designs retain runtime commitment within established controllers.}
\label{fig:era3-patterns}
\end{figure*}

Together, these architectural and temporal considerations place the boundary between proposal generation and authorized execution at the center of LLM-enabled network operations. According to our study, available evidence does not yet support assigning LLM-enabled agents direct, end-to-end authority over operational networks. Current designs generally take a conservative approach: LLM-enabled components interpret requests, assemble evidence, and propose actions, while approval and execution remain subject to established operational controls. Broader autonomy requires evidence that proposed changes have been validated, execution remains within authorized constraints, outcomes can be monitored, and failures can be contained or reversed~\cite{krentsel2026crosscheck,renganathan2023hydra,xiang2023tulkun,zhao2024epverifier,li2025ndd,wang2025s2,yang2026s2sim,yuan2025hoyan,schneider2025velo}. The remainder of this section first derives operator requirements from this authority boundary, then develops a reference architecture, identifies recurring patterns for integrating LLM-enabled components into network operations, and analyzes the emerging trend through the NCI framework.

\subsection{Operator Requirements}
\label{sec:eraIII-req}

From the operator's perspective, LLM-enabled automation is operationally acceptable only when it is integrated into controlled OSS/OAM processes. It cannot be deployed as an unconstrained control agent. Four requirements follow. First, proposal generation must remain separate from commit authority. Second, proposed actions must be represented in forms that can be checked against schemas, policies, safety invariants, reachability constraints, and the data models exposed by control and management systems. Third, execution must support controlled deployment, post-change monitoring, and recovery when outcomes deviate from expectations. Fourth, the workflow must produce auditable records that link each request to its inputs, validation results, executed actions, and observed outcomes. These properties must remain effective across multiple vendors, integrate with existing Operations Support Systems and Network Management Systems (OSS/NMS), and fail safely when telemetry is incomplete, tools are unavailable, or operator intent remains ambiguous~\cite{IG1251E,etsi-gs-zsm}.

\subsection{Reference Architecture}
\label{sec:eraIII-refarch}

Fig.~\ref{fig:era3-refarch} presents a minimal reference architecture organized around the boundary between proposal generation and authorized execution. It does not replace complete OSS/NMS, Zero-touch Network and Service Management (ZSM), or autonomous-network architectures. Instead, it isolates the functions needed to produce, validate, authorize, execute, and record an operational change.

On the proposal side of the boundary, an interaction and proposal layer interprets operator intent and combines it with logs, topology state, tickets, configuration information, and other operational evidence. It produces inspectable outputs such as diagnostic hypotheses, verification queries, proposed configuration changes, or rollout plans. Before execution, a governed validation pipeline checks whether the proposal satisfies applicable schemas, policies, invariants, authorization rules, and safety constraints. Approved proposals are then applied through established controllers, orchestrators, or management systems~\cite{reitblatt2012abstractions,enns2011network,bierman2018network, IG1251E}.
A knowledge and observability plane provides the operational information needed for proposal generation and validation, including logs, metrics, traces, topology, inventory, policy context, and digital twin state. Simulation and digital-twin facilities can additionally support what-if evaluation of proposed changes~\cite{almasan2022network,wang2026arcadia}. An assurance and audit plane preserves provenance, validation results, authorization decisions, deployment observations, rollback triggers, and finalized change records.

Operationally, a request and its supporting evidence are assembled into a proposal artifact. When the proposal is intended to produce an operational change, it enters the governed validation pipeline. Proposals that fail schema, policy, invariant, or authorization checks are returned to the proposal layer with validation feedback instead of being executed. Approved proposals are applied through existing controller interfaces and management protocols, deployed in controlled stages when appropriate, and monitored for conditions that require rollback or other recovery actions. The workflow concludes by producing a \emph{change record} that links the original request, supporting evidence, validation results, authorization decision, executed actions, and observed outcomes~\cite{wang2024netassistant,wang2025intent,mekrache2025oss,reitblatt2012abstractions,bierman2018network,krentsel2026crosscheck,yang2026s2sim,yuan2025hoyan,nelson2025incident}.

\subsection{Integration Patterns}
\label{sec:eraIII-patterns}
This subsection classifies representative systems in the LLM-enabled era by their operational role and authority boundary. Our study identifies three recurring integration patterns, namely \emph{operator-facing assistants}, \emph{tool-using OAM workflow agents}, and \emph{controller-centered execution with LLM-generated objectives}. Assistants primarily improve interaction and proposal generation; workflow agents extend the LLM-enabled component to coordinate multi-step procedures via approved tools; and controller-centered designs use LLMs to express objectives or constraints, while established controllers retain runtime execution authority.
The patterns differ in the role assigned to the LLM-enabled component, the artifacts or procedures it produces, the boundary at which human approval or controller-mediated execution is required, and the assurance mechanisms that constrain any proposed or executed change. 
Fig.~\ref{fig:era3-patterns} places these three patterns along a common operational pipeline and highlights where LLM-supported processing terminates relative to the commit-authority boundary.


\subsubsection*{Operator-facing assistants}
In this pattern, the assistant serves as a bridge between network operators and existing OAM tools. It helps translate an operator request into a relevant operational context. This context may include operational records and network state, such as logs, tickets, runbooks, topology information, and configuration data. Based on this context, the assistant may identify relevant procedures and produce diagnostic hypotheses, checklists, or draft configuration changes. When such outputs may affect network state, they remain subject to operator review and to the approval processes of existing control and management systems.

One important category is the diagnostic assistant, which uses LLMs to help operators interpret incidents and localize likely causes. NetAssistant~\cite{wang2024netassistant} is a production-reported assistant for data-center diagnosis. It uses an LLM-backed chatbot to translate operator questions into predefined diagnostic workflows that query monitoring data and return diagnostic results. BiAn~\cite{wang2025towards} is deployed in Alibaba Cloud's production network for failure localization. It organizes the localization process through LLM-assisted agents that analyze monitoring data, score candidate devices, and produce ranked failure-localization results with explanations. RCACopilot~\cite{chen2024automatic}, although focused on cloud incidents rather than network-specific OAM, uses LLM-based reasoning over collected diagnostic information to predict a root-cause category and produce an explanatory report for human responders. In these diagnostic assistants, the LLM supports interpretation, diagnosis, and explanation; yet operational action remains with human operators and the surrounding control or management processes.

A second assistant category focuses on configuration and test authoring. Unlike diagnostic assistants, which primarily support investigation, these systems generate artifacts that may later enter validation, testing, or deployment workflows. NetConfEval~\cite{wang2024netconfeval} benchmarks natural-language-to-configuration and related network-programming tasks. Clarify~\cite{mondal2025tackling}, CEGS~\cite{liu2025cegs}, and INTA~\cite{wei2025inta} provide research prototypes for ambiguity resolution, documentation-grounded configuration synthesis, and cross-vendor configuration translation with syntax or semantic checks. LLM-NetCFG~\cite{llmnetcfg} illustrates an intent-to-configuration workflow that combines local LLM-based generation with verification and device configuration. EYWA~\cite{mondal2026eywa} extends the same authoring role to protocol testing. It employs LLMs to construct intended protocol-behavior models from RFCs and related documents, after which model-based testing exercises implementations. In these assistants, the LLM-generated artifact is closer to operational action than the diagnostic ones, but its use still depends on validation and the surrounding execution process.

From the NCI perspective, assistants usually reach \emph{D2}, and some approach \emph{D3} when they construct or revise multi-step plans at the proposal layer. Their behavior remains configured or curated rather than self-updating (\emph{A0--A1}). Knowledge ranges from \emph{K2} to \emph{K3}, depending on whether the system primarily uses structured documentation and task inputs or correlates heterogeneous operational records. Interface capability similarly ranges from intent translation to dialogue clarification (\emph{I2--I3}). Control Delegation remains low (\emph{C0--C1}) because commit authority stays outside the assistant.

\subsubsection*{Tool-using OAM workflow agents}
This pattern extends the assistant role from producing reviewable outputs to coordinating approved operational procedures. The agent receives an operator request, breaks it into ordered steps, invokes relevant tools, and records intermediate results. The tools may retrieve information, run diagnostics, update tickets, or perform operations that have already been authorized by the surrounding process. The agent thus coordinates execution within a bounded workflow, but it does not grant itself authority to change the network. Operations that may affect network state remain subject to the permissions, checks, and approval processes of existing control and management systems.

Existing workflow agents span production settings and research prototypes. Confucius~\cite{wang2025intent} reports two years of production operation with more than 60 onboarded applications. It converts operator intent into workflow graphs that reuse existing operational procedures and runbook tools. Before an operation that may change the network state is executed, the workflow is verified and validated. OSS-GPT~\cite{mekrache2025oss} presents an experimental design for coordinating operations across support-system interfaces. The agent plans a sequence of API calls, but operations that create, update, or delete service state require human validation. ArachNet~\cite{ramanathan2025towards} applies a similar tool-using workflow to Internet measurement tasks. Its prototype implementation coordinates multiple tools and records intermediate results.

\begin{figure*}[!tb]
\centering
\includegraphics[width=1\textwidth]{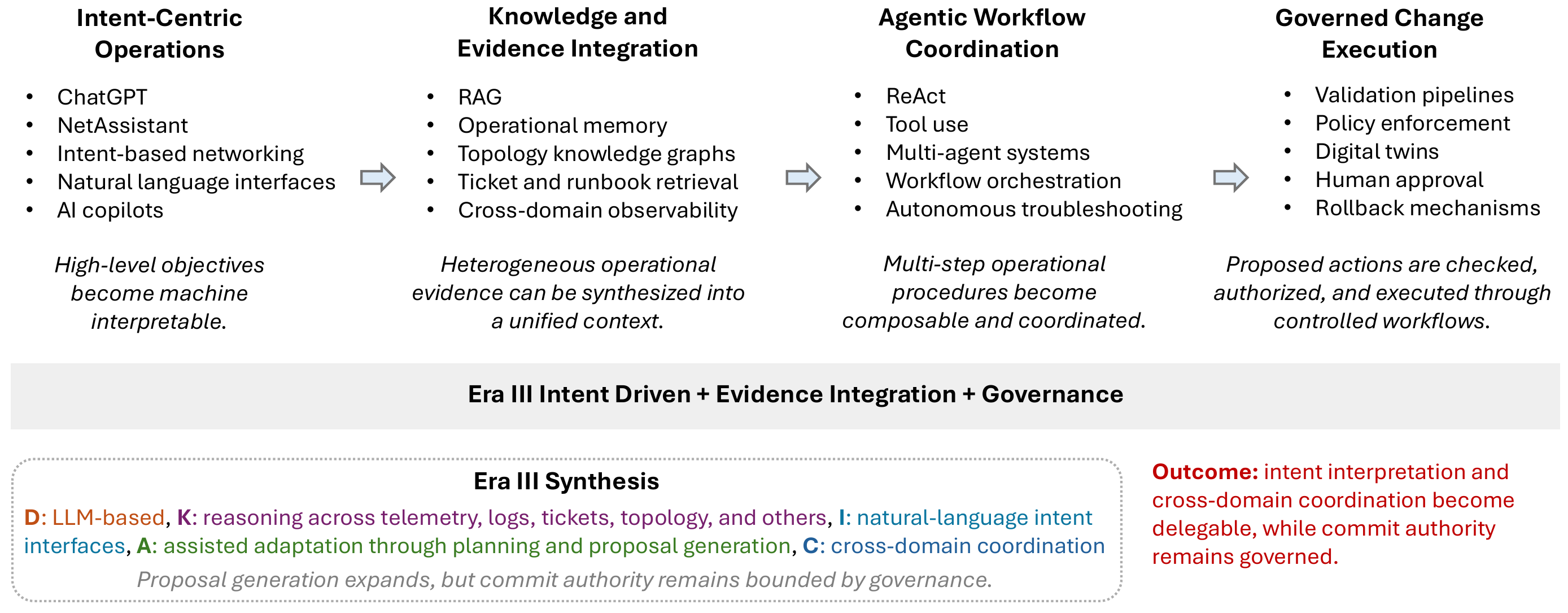}
\caption{This era transitions toward LLM-enabled network operations. Cross-domain operational pressure from network slicing, cloud-native functions, disaggregated RAN/core deployments, and multi-vendor environments shifts the bottleneck from domain-local control toward intent interpretation, evidence synthesis, and workflow coordination. LLM-enabled components operate mainly on the proposal side: they retrieve operational context, synthesize evidence, and generate proposal artifacts such as diagnostic hypotheses, verification queries, configuration deltas, ticket updates, rollout plans, and objective specifications. Governed execution remains separated from proposal generation through pre-change checks, authorization, controller-mediated execution, monitoring, rollback, and audit records. Through the NCI lens, Era III advances \emph{Interface} and \emph{Knowledge} most strongly, broadens proposal-side \emph{Decision Logic}, and leaves \emph{Control Delegation} and \emph{Adaptability} bounded by assurance requirements.}
\label{fig:era3-llm-enabled-operations}
\end{figure*}

In the NCI profile, tool-using workflow agents mainly advance \emph{Decision Logic}. They can reach \emph{D3} when the agent constructs, compares, or revises a multi-step procedure; systems that only select a predefined workflow remain closer to \emph{D1--D2}. Their \emph{Adaptability} is usually limited to \emph{A1} because most systems reuse existing procedures rather than updating models, policies, or configurations in response to operational feedback. Their \emph{Knowledge} can reach \emph{K3} when the workflow combines operational records, documentation, procedure repositories, and execution traces. Their \emph{Interface} capability is typically \emph{I2--I3}, because the agent translates operator requests into tool calls and returns intermediate results in a form operators can inspect. \emph{Control Delegation} rises to \emph{C1--C2}, since the agent may execute approved workflows, but commit authority remains outside the agent.

\begin{table*}[t]
\centering
\scriptsize
\caption{Five-axis profile of representative works in the LLM-enabled era. The table highlights a recurring asymmetry: \emph{Interface} and \emph{Knowledge} advance quickly, while \emph{Control Delegation} remains conservative. Adaptability entries above \emph{A1} are reserved for periodic data-driven updates or continuous online adaptation, consistent with Section~\ref{sec:framework}; ordinary maintenance of prompts, retrieval corpora, workflows, or policy rules does not, by itself, qualify. The evidence column distinguishes production reports, research prototypes, and benchmark studies.}
\label{tab:era3-fiveaxis}
\begin{tabular}{p{3.2cm}ccccc p{2.4cm}}
\toprule
\textbf{Work / System} &
\textbf{D} &
\textbf{A} &
\textbf{K} &
\textbf{C} &
\textbf{I} &
\textbf{Evidence} \\
\midrule

\multicolumn{7}{l}{\textbf{Operator-facing assistants: diagnosis}}\\
NetAssistant~\cite{wang2024netassistant} & D2 & A1 & K3 & C0 & I3 & Production report \\
BiAn~\cite{wang2025towards} & D2 & A1 & K3 & C0 & I3 & Production report \\
RCACopilot~\cite{chen2024automatic} & D2 & A1 & K3 & C0 & I2 & Research prototype \\
\midrule

\multicolumn{7}{l}{\textbf{Operator-facing assistants: configuration and test authoring}}\\
NetConfEval~\cite{wang2024netconfeval} & D2 & A0 & K2 & C0 & I2 & Benchmark study \\
Clarify~\cite{mondal2025tackling} & D2 & A1 & K2--K3 & C1 & I3 & Research prototype \\
CEGS~\cite{liu2025cegs} & D2 & A1 & K2--K3 & C1 & I2 & Research prototype \\
INTA~\cite{wei2025inta} & D2 & A1 & K2--K3 & C1 & I2 & Research prototype \\
LLM-NetCFG~\cite{llmnetcfg} & D2 & A1 & K2--K3 & C1 & I2 & Research prototype \\
EYWA~\cite{mondal2026eywa} & D2 & A0 & K3 & C0 & I2 & Research prototype \\
\midrule

\multicolumn{7}{l}{\textbf{Tool-using OAM workflow agents}}\\
Confucius~\cite{wang2025intent} & D3 & A1 & K3 & C2 & I2 & Production report \\
OSS-GPT~\cite{mekrache2025oss} & D3 & A1 & K3 & C1 & I3 & Research prototype \\
ArachNet~\cite{ramanathan2025towards} & D3 & A1 & K3 & C1 & I2 & Research prototype \\
\midrule

\multicolumn{7}{l}{\textbf{Controller-centered execution with LLM-generated objectives}}\\
IntentOpt~\cite{ahmed2026vision} & D2 & A0 & K2--K3 & C0 & I2 & Benchmark study \\
LLM--solver allocation~\cite{sudhakara2025constrained} & D2 & A0 & K2 & C0 & I1--I2 & Research prototype \\
LeJIT~\cite{he2025just} & D2 & A0 & K2 & C0 & I1 & Research prototype \\
\midrule

\multicolumn{7}{l}{\textbf{Controller-centered systems with limited or no LLM role}}\\
NetNomos~\cite{he2026netnomos} & D2 & A0 & K2--K3 & C0 & I1 & Research prototype \\
Matryoshka~\cite{cai2026matryoshka} & D1--D2 & A1 & K2 & C1--C2 & I2 & Production report \\
NetKeeper~\cite{wan2025netkeeper} & D2--D3 & A2 & K2--K3 & C1--C2 & I2 & Research prototype \\
\bottomrule
\end{tabular}
\end{table*}

\subsubsection*{Controller-centered execution}
This pattern keeps execution inside established control mechanisms, such as controllers, solvers, and compilers. The LLM does not run the control loop directly. Instead, it translates operator intent, service requirements, or scenario descriptions into structured inputs, such as objectives, constraints, parameters, or test cases, that these mechanisms can process. Existing validation and approval processes check the resulting specification before any network change is applied. Runtime execution remains with controllers, solvers, and compilers, while validators and approval processes determine whether the proposed change can be applied.

Most systems in this pattern are either benchmarks or research prototypes. For instance, IntentOpt~\cite{ahmed2026vision} evaluates whether LLM-generated optimization formulations match the requested objective and can be executed against ground-truth. It shows that textual plausibility alone is not enough; a formulation must also be compilable, feasible, and consistent with the requested behavior. Hybrid LLM-solver designs for constrained allocation~\cite{sudhakara2025constrained} and solver-guided generation in LeJIT~\cite{he2025just} follow the same division of labor. The LLM prepares a structured formulation, while solvers or compilers verify its validity and enforce the applicable constraints.

The same boundary also manifests in controller-centered systems, where the LLM is not the main source of the objective. Matryoshka~\cite{cai2026matryoshka} is a production-scale, model-driven compiler that translates high-level data-center design intent into detailed switch configurations. NetKeeper~\cite{wan2025netkeeper} combines natural-language intent, anomaly information, and traffic analysis with RL-based configuration updates in dynamic networks. NetNomos~\cite{he2026netnomos} uses formal rules and solver-based checks to constrain generative outputs. Network digital twins and high-fidelity simulation platforms provide another form of boundary: they allow proposed objectives or configuration updates to be tested before deployment~\cite{almasan2022network,wang2026arcadia}. These works broaden the pattern beyond LLM-only formulation, but they support the same architectural point.

In the NCI view, this pattern mainly affects \emph{Interface} and \emph{Decision Logic}. The \emph{Interface} score typically ranges from \emph{I1--I2} because the system translates operator intent into structured objectives or constraints. \emph{Decision Logic} is usually \emph{D2}, as the LLM helps formulate the decision problem, but a solver, compiler, controller, or simulator makes the control decision. \emph{Adaptability} often remains \emph{A0--A1}, because the underlying control logic changes slowly and is not usually updated by the LLM itself. \emph{Knowledge} can reach \emph{K3} when the generated specification is based on topology, traffic, policies, or simulation state. \emph{Control Delegation} remains low, typically \emph{C0--C1}, because runtime authority stays with the established control mechanism rather than the LLM. Model-driven compilers and adaptive update engines may move further along \emph{Control Delegation}, but their authority comes from explicit models, validators, and approval processes rather than unconstrained generative reasoning.

\subsubsection*{Capabilities and Boundaries}
Across the three integration patterns, LLMs mainly strengthen the proposal side of network operations. They help interpret operator requests, assemble relevant operational context, draft reviewable outputs, and coordinate the steps of approved procedures. In doing so, they can make logs, tickets, runbooks, topology information, and configuration data easier to use in operational workflows. Their intermediate steps and outputs should remain visible to operators and existing management systems, rather than being hidden inside the model.

The same evidence also clarifies that LLMs do not remove the need for domain controllers, optimization engines, policy checks, validation mechanisms, approval processes, or recovery procedures. When a system appears autonomous, the autonomy is usually assembled from several parts: LLM-assisted proposal generation, execution through established controllers or tools, and explicit checks before a network change is applied. The scope of action remains limited by permissions and by the surrounding operational process. Fig.~\ref{fig:era3-llm-enabled-operations} summarizes the shift toward LLM-assisted proposal generation, evidence synthesis, and workflow coordination, with operational changes applied through governed execution paths.

The evidence behind these patterns is uneven. Production deployments are strongest for diagnostic assistants and tool-using OAM workflow agents, as illustrated by NetAssistant and Confucius. Matryoshka and Hoyan show that intent compilation and verification can become production-grade operational infrastructure, even when the LLM role is limited~\cite{cai2026matryoshka,yuan2025hoyan}. By contrast, controller-centered uses of LLMs are still dominated by benchmarks and research prototypes, such as IntentOpt, LeJIT, and constrained LLM-solver designs.

\subsection{Five-Axis Analysis of the LLM-Enabled Era}
\label{sec:eraIII-fiveaxis}
The integration patterns above indicate that LLM-enabled components mostly enter operational workflows before approval and execution, as summarized in Table~\ref{tab:era3-fiveaxis}. The most visible movement is along the \emph{Interface} and \emph{Knowledge} axes. LLM-enabled systems reduce the effort required to turn operator requests, operational records, runbooks, topology information, and configuration data into explanations, reviewable proposals, and structured inputs that existing controllers can process. Assistants and workflow agents most directly advance \emph{Interface} and \emph{Knowledge}. They also broaden \emph{Decision Logic} at the proposal layer while keeping \emph{Control Delegation} low, since approval and execution are still tied to established processes.

The five-axis profile exposes the main asymmetry of this era: LLM-enabled systems have become better at interpreting requests, assembling evidence, and drafting operational proposals, but this progress has not been matched by a comparable increase in their authority to implement changes or their ability to adapt safely during operation. Richer interfaces allow operators to express intent more naturally, but they also require clarification, grounding, and validation when intent is ambiguous. Broader operational knowledge improves proposal quality only when the source, freshness, scope, and provenance of the supporting context are visible. Likewise, stronger proposal-side decision logic can improve diagnosis, change authoring, and workflow coordination, but these gains do not by themselves justify greater authority to apply changes.

By contrast, \emph{Control Delegation} and \emph{Adaptability} are still the slower-moving axes. Although tool-using workflow agents move further along \emph{Control Delegation} than assistants because they can execute approved procedural steps using existing tools, the authority to commit still lies outside the LLM-enabled component. Controller-centered designs are more conservative: the LLM prepares objectives, constraints, or structured inputs, while controllers, solvers, or compilers retain runtime authority. \emph{Adaptability} is also limited in most systems. Operational behavior is usually changed through maintained prompts, retrieval sources, workflows, model versions, policy rules, configuration templates, or controller parameters, not through online adaptation by the LLM during operation.

This asymmetry suggests a cautious view of the LLM-enabled era. Its main contribution is not autonomous actuation, but the generation of proposals and workflows that can be inspected, checked, approved, and recorded. Consequently, evaluation should reflect the authority assigned to the LLM-enabled component. A proposal-generating assistant should be judged by whether its outputs are grounded, reviewable, and checkable; a workflow agent that approaches state-changing operations must also be judged by whether permissions, approval gates, monitoring, rollback, and audit records work as intended. Task success and linguistic plausibility are useful signals, but they are insufficient for assessing operational control.
The next step is not to expand LLM autonomy simply because proposal generation has improved, but to pragmatically establish when proposal generation, controlled adaptation, and authorized execution are supported by sufficient evidence, safeguards, and accountability within the surrounding operational process, as detailed in Section~\ref{sec:challenges}.


\section{Toward Trustworthy Autonomous Networking}
\label{sec:challenges}

The evolution traced in Sections~\ref{sec:era-I}--\ref{sec:era-III} shows the uneven progress of NCI. Specifically, rule-based automation made routine operations more repeatable; programmable and data-driven control improved observability, optimization, and domain-specific control; and LLM-enabled systems now strengthen intent interpretation, evidence synthesis, and proposal generation. However, these advances do not by themselves justify broader network autonomy. A network control system that can observe more states, generate better explanations, or prepare more structured proposals may still be unsuitable for changing network state without additional checks, authority boundaries, and recovery mechanisms.

This section translates the NCI lens into a forward-looking roadmap for trustworthy autonomy in evolving network environments. The roadmap focuses on two coupled questions. First, can NCI remain reliable as traffic, topology, software, policies, and operational practices change? Second, can autonomy increase without weakening the safeguards that make network changes accountable? The TM Forum Autonomous Networks Levels framework describes staged progress toward more autonomous operation, largely in terms of how operational tasks and authority are delegated between humans and automated systems~\cite{ig1392}. Recent industry reports show that operators are actively pursuing this direction~\cite{tmforum-regional-an-progress}. The NCI perspective asks a complementary question: what technical conditions must be in place before such autonomy becomes defensible? In practice, this remains difficult when integration, data governance, regulatory constraints, skills, organizational trust, and cross-domain assurance are not yet mature~\cite{capegemini-report}. For this reason, stronger autonomy claims should be tied to explicit scope, authority boundaries, operational state, validation procedures, and recovery conditions~\cite{appledore-autonomy-chasm}.

On this basis, we organize the roadmap around \emph{Adaptability} and \emph{Control Delegation}. Adaptability determines whether NCI can remain reliable as operating conditions change. Control Delegation determines whether authority to affect network state can be delegated to automated components without weakening accountability. The other three axes operationalize this focus: \emph{Knowledge} grounds decisions in a faithful view of the network state, \emph{Decision Logic} makes candidate actions checkable under constraints, and \emph{Interface} exposes objectives, evidence, uncertainty, and responsibility to operators and peer systems. Cross-domain incidents and changes illustrate why these axes must be considered together, as symptoms and dependencies may span services, topology, policies, telemetry, recent configuration changes, and rollout constraints~\cite{etsi-zsm-009,etsi-zsm-016,almasan2022network,clemm2022intent}. Trustworthy autonomous networking thus requires that the five axes mature together and be evaluated under realistic operational conditions, including pre-change validation, incident replay, rollback behavior, and production-level workflows~\cite{brown2023batfish,gao2024crescent,sentosa2025cellreplay}.

\subsection{Evaluation Methodology and Evidence of Progress}
\label{subsec:challenge-evaluation}
Before delving into the remaining technical challenges, it is necessary to clarify what would count as {\em evidence of progress}. In autonomous networking, benchmark accuracy, successful troubleshooting on curated tickets, or a plausible natural-language explanation can be useful, but they do not prove that a system can be safely granted more authority over network-side actuation. For instance, the systems reviewed in Section~\ref{sec:era-III} are usually hybrid, i.e., LLM-enabled components interpret requests, assemble evidence, and prepare proposals, while controllers, solvers, validation tools, approval processes, and recovery procedures remain responsible for execution. The evaluation should state which part of this workflow is being tested and where the authority boundary is drawn.

To make evidence comparable, each study should specify the NCI capability being tested and the authority boundary assumed by the system. Specifically, it should state whether the system produces only advisory outputs, coordinates approved steps, or can affect the network state; it should also identify the associated checks, approvals, guardrails, and recovery mechanisms. When the system adapts its models, workflows, policies, or configuration over time, the evaluation should show whether the original evidence remains valid or whether the system must be revalidated before stronger autonomy is bestowed. Without this separation, a system that improves diagnosis may be mistaken for one that can execute remediation, and a system that coordinates approved procedure steps may be mistaken for one that holds independent commit authority. Evaluation should report both effectiveness and efficiency: whether actions remain valid and recoverable under change, and whether the control process stays within acceptable purviews for latency, cost, and operator workload.

\subsection{Knowledge: Faithful Operational State}
\label{subsec:challenge-knowledge}
The first axis-specific bottleneck is \emph{Knowledge}. Autonomous operation depends on the state over which the system reasons. In many deployments, telemetry, topology, configuration, inventory, service dependencies, tickets, and policy artifacts are distributed across tools, domains, vendors, and administrative domains~\cite{almasan2022network,clark2003knowledge,etsi_zsm015_2024}. Even when each domain is well instrumented, cross-domain context may be reconstructed from records with different timestamps, naming conventions, levels of detail, and provenance. An autonomous system can then appear competent on abundant local evidence while failing when the relevant cause spans domains.

Moreover, trustworthy autonomy requires more than collecting more data. It requires an operational knowledge base that represents what the network is doing, what it is expected to do, what constraints it must satisfy, and how reliable each piece of evidence is. The system must also track the conditions under which each piece of context is valid. A topology snapshot, ticket, policy rule, or telemetry stream may be reliable only within a particular time window, domain, or administrative scope. When those limits are unclear, an otherwise valid proposal can rely on stale evidence, misattribute responsibility, or apply a change in the wrong part of the network.
Digital twins illustrate both the promise and the difficulty. They can support what-if analysis, pre-change validation, and cross-domain reasoning, but only if their assumptions and synchronization state are explicit~\cite{almasan2022network,wu2021digital,etsi-zsm-018}. A twin used for planning without operational reality may provide a false sense of reliability. Similarly, knowledge graphs and schema-driven pipelines can encode topology, policies, service dependencies, and historical incidents, but they must remain accurate through failures, maintenance windows, and configuration changes~\cite{clemm2022intent,ujcich2020provenance}. The research challenge is to build knowledge layers that can be checked against the live network and invalidated when their assumptions no longer hold.

Evidence on this axis should demonstrate that the system can reliably reconstruct the operational state under realistic conditions. This can be evaluated through replay studies on historical incidents, shadow-mode studies beside live operations, or controlled fault-injection experiments in emulation or digital-twin environments. Such studies should compare inferred state with trusted operational records, measure the impact of stale or missing inputs, and test whether incorrect context is detected before it affects a proposal or a network change. Without such evidence, stronger decision logic or more capable language models will be poorly grounded.

\subsection{Decision Logic: Compositional Planning Under Operational Constraints}
\label{subsec:challenge-decision}
Once the system has a usable view of operational state, the next question is how it turns that state into valid actions. The \emph{Decision Logic} challenge is not to replace existing control mechanisms with a general-purpose reasoner. Multi-domain production networks already rely on specialized mechanisms, including traffic-engineering solvers, admission-control logic, configuration compilers, verification tools, diagnosis systems, and domain-specific control loops. The challenge is to compose these mechanisms so that plans remain valid across domains, timescales, and approval and execution boundaries.

Section~\ref{sec:era-III} suggests a recurring division of labor. LLM-enabled components are useful for interpreting requests, assembling context, drafting proposals, constructing structured formulations, and coordinating steps in approved procedures. Actions that require responsiveness, stability, or safety guarantees should remain with mechanisms whose behavior can be bounded and checked. In controller-centered systems, for example, an LLM may prepare objectives or constraints, while solvers, compilers, and controllers determine whether the resulting specification is valid and executable~\cite{ahmed2026vision,he2025just}. In tool-using workflow agents, the agent may sequence approved steps, but authority to apply network changes remains with the surrounding control and approval process~\cite{wang2025intent,mekrache2025oss}.

The open problem is to make the assumptions behind each proposed action explicit. A proposal should state the conditions under which it is valid, including required inputs, preconditions, affected scope, expected outcomes, rollback conditions, and checks to be performed before execution. The same requirement applies to plans assembled from multiple tools. A sequence that is safe within one domain may become unsafe once service dependencies, maintenance windows, or policy constraints in another domain are taken into account. Decision logic should produce artifacts that can be checked, simulated, and audited in addition to explanations.

Future work should make the handoff from proposal generation to governed execution explicit. A proposal artifact should state the information it relies on, the intended scope of the change, the checks it has passed, the assumptions that still need validation, and the conditions under which rollback would be triggered. This is necessary because a change that appears safe within one domain may become unsafe when service dependencies, policy rules, maintenance windows, or shared resources in another domain are considered~\cite{renganathan2023hydra,guo2022flash,xu2024relational}. Clear handoff records allow existing control and management systems to check a proposal before execution, monitor whether its assumptions hold after deployment, and recover if the outcome deviates from the expected path.

\subsection{Adaptability: Improving Without Destabilizing}
\label{subsec:challenge-adaptability}

As for \emph{Adaptability}, the challenge is no longer simply whether a system can adjust its behavior. The harder question is whether such adjustments can be introduced without weakening the guarantees under which the system was allowed to operate. In autonomous networking, control behaviors can be adapted through several sources, including prompts that guide LLM responses, retrieval sources that supply operational context, workflow definitions that order tool use, policy rules that constrain approval, learned models that support prediction or diagnosis, configuration templates that generate candidate changes, and controller parameters that shape runtime decisions.
Each adaptation can improve performance in one setting while creating regressions elsewhere, moving behavior outside tested envelopes, or degrading performance on unseen but operationally important cases~\cite{kirkpatrick2017overcoming,wachi2024saferl}.

This distinction is important when interpreting the \emph{Adaptability} axis. Routine maintenance of prompts, retrieval corpora, workflow definitions, or policy files should not be treated as high adaptability. Such maintenance may improve system behavior, but it does not necessarily show that the system learns from operational feedback or updates its behavior through a governed adaptation process. Higher levels on the Adaptability axis should be reserved for systems that periodically or continuously update models, policies, parameters, or configurations based on operational data, and that subject those updates to regression control and human or automated oversight.

This distinction also exposes several failure modes. Learning-enabled controllers can perform well on fixed topologies or curated fault scenarios, yet degrade under traffic shifts, new hardware, or unseen incidents~\cite{bernardez2021machine,gui2024redte,cai2023deep}. LLM-enabled workflows can drift over long tasks when intermediate state is summarized poorly, or tool results are not preserved in durable artifacts~\cite{anthropic2025context,openai2026harness}. Configuration-update systems can improve one metric while violating an operational constraint that was absent from the training or update loop. In all three cases, the problem is not adaptation itself, but adaptation without adequate control over regression, scope, and recovery.

Evidence should be reported for the specific artifact or component that has been updated. For a learned model, this includes performance under shifted topologies, traffic patterns, faults, and policy settings, as well as regression results on previously validated cases. For a workflow, policy rule, prompt package, retrieval corpus, or configuration template, this includes tests showing that existing procedures still execute correctly, unsafe proposals are still blocked, required checks are still applied, and operational records remain complete. Across these cases, the evaluation should report how failures are detected, whether rollback succeeds, how long recovery takes, and whether the system safely reduces its autonomy level until the update is corrected and revalidated. Adaptation that cannot preserve or safely lower autonomy is not operational progress; it is an unbounded change to the assurance envelope.

\begin{figure*}
\includegraphics[width=\textwidth]{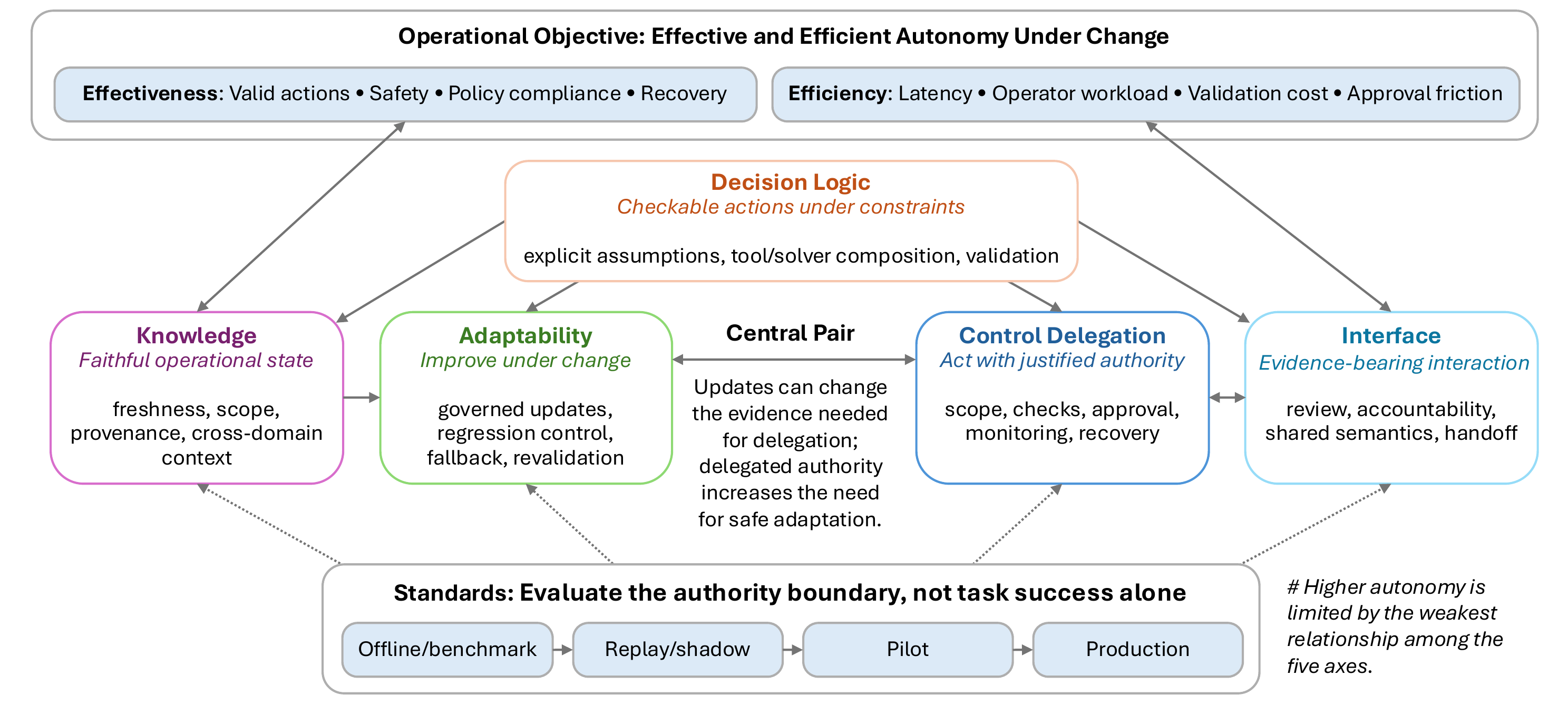}
\caption{Roadmap toward trustworthy autonomous networking through the five-axis NCI lens. \emph{Adaptability} and \emph{Control Delegation} form the central unresolved pair: a system must improve under changing conditions, and stronger authority must remain justified after such change. \emph{Knowledge}, \emph{Decision Logic}, and \emph{Interface} provide the supporting conditions. Evidence standards determine whether stronger autonomy claims are credible.}
\label{fig:trustworthy-roadmap}
\end{figure*}

\subsection{Control Delegation: Authority as an Assurance Claim}
\label{subsec:challenge-delegation}

\emph{Control Delegation} is the second central bottleneck because it determines when a system can move from proposal to action. Broader autonomy is not justified by better recommendations alone, and it becomes especially fragile when the system is also adapting. A proposed change must remain within scope, satisfy policy and safety constraints, be rolled out safely, be monitored after deployment, and be recoverable if the outcome is unacceptable~\cite{etsi-zsm-016,etsi-zsm-009,brown2023batfish,renganathan2023hydra}.

In many systems, these checks remain fragmented. For instance, a configuration may be syntactically valid but inconsistent with policy. A routing update may satisfy a local objective but violate a service-level constraint elsewhere. A workflow may include approval steps but leave incomplete evidence about what was checked, who or what approved the action, and why execution was allowed to proceed. Stronger autonomy requires these checks to be tied to both the proposal artifact and the execution record.

A delegated action should leave enough evidence for operators to understand why execution was allowed. The record should connect the original intent to the affected resources, the checks that constrained the action, the approval decision, the rollout strategy, and the conditions under which recovery would be triggered. This evidence can be produced through verification, staged rollout, shadow execution, production emulation, and replay~\cite{fogel2015batfish,beckett2017minesweeper,zhao2024epverifier,li2025ndd,almasan2022network,gao2024crescent,sentosa2025cellreplay}. As autonomy increases, the record must provide stronger justification that the action remained within its declared scope.

Security and governance also determine how far control can be delegated. Tool-using LLM systems expand the attack surface because retrieved content, tool outputs, and intermediate instructions can influence later actions. Indirect prompt injection, excessive tool permissions, and misuse of operational data are not only security concerns; they can invalidate the authority boundary itself~\cite{zhan2024injecagent}. Telecom-oriented agent protocols add another layer of coordination, where capabilities, permitted scope, and accountability must be explicit rather than inferred~\cite{IG1453A2AT}. A defensible design should restrict what each tool can do, preserve the origin of retrieved evidence, isolate operations that can change state, and leave an audit trail from request to outcome~\cite{ujcich2020provenance,nist2020zta}. These controls determine whether a system can be trusted to apply changes, not merely whether it is securely implemented.

\subsection{Interface: Evidence-Bearing Interaction}
\label{subsec:challenge-interface}

The \emph{Interface} axis determines how objectives, explanations, decisions, and supporting records are exchanged. The LLM-enabled era makes this axis central because natural-language interaction can reduce the effort required to express intent, inspect incidents, and coordinate operations. Yet conversational fluency does not provide accountability by itself. An interface that supports stronger autonomy must make the basis for a recommendation or action visible: which information was used, which assumptions were made, which checks passed or failed, and what uncertainty remains~\cite{IG1251E,IG1414,etsi-zsm-016}.

For operators, the interface should evolve from a place for issuing commands into a workspace for reviewing evidence and responsibility. It should connect the original request, the assembled context, the proposal artifact, the validation results, the approval decision, and the observed outcome. This requirement follows directly from the proposal/execution boundary in Section~\ref{sec:era-III}. When an LLM-enabled component proposes a change, the interface should make the proposal and its supporting evidence reviewable before the change is applied. When a workflow agent executes approved steps, the interface should expose the execution trace, deviations from the expected path, and any escalation.

Interoperability extends the same problem across tools and organizations. Automation components should describe their capabilities, required inputs, permitted actions, and safety scope in forms that other systems can interpret. Intent and state should be exchanged with stable semantics across integration points. Conflict handling, arbitration, and plan repair should be explicit parts of the interface~\cite{clemm2022intent, IG1453A2AT,mehmood2023intent,biyar2025autonomous}. Without shared semantics, each new integration must rebuild the assurance boundary from the beginning.

The human role also changes. Operators move from writing low-level commands toward reviewing proposals, curating policies, handling exceptions, and learning from incidents. This transition requires training and organizational processes that define responsibility for review, approval, and escalation. A system may be technically capable of producing valid proposals, but autonomy will remain limited if operators cannot understand why an action was proposed, what risk remains, and who is accountable for approving it.

\subsection*{From Axes to Roadmap}
\label{subsec:challenge-synthesis}

The challenges across the five axes are not a checklist of five independent workstreams. The roadmap is illustrated in Fig.~\ref{fig:trustworthy-roadmap}. Its center is the Adaptability--Control Delegation pair, as a network control system must be able to improve as conditions change, and any increase in authority must remain justified after that change. The other three axes provide the supporting conditions: Knowledge keeps adaptation and autonomy grounded in a faithful state; Decision Logic makes proposed actions checkable; and Interface makes evidence, uncertainty, and responsibility visible. Higher autonomy is limited by the weakest relationship among these axes, not by the most advanced component.

This roadmap also clarifies why the research--deployment gap persists. Emulation, replay, configuration analysis, and deployed diagnostic systems show that production environments impose constraints on scale, fidelity, reliability, and operational workflow that curated laboratory studies can miss~\cite{gao2024crescent,sentosa2025cellreplay,brown2023batfish,wang2024netassistant}. Credible progress should be demonstrated under realistic constraints, including change windows, rollback requirements, multi-vendor boundaries, partial observability, and adversarial conditions. The relevant technologies should be evaluated as parts of an operational control stack rather than as isolated advances. In this sense, the roadmap is not a list of desirable features; it is an evidentiary discipline for deciding when a capability can safely support stronger autonomy.


\section{Conclusion}
\label{sec:conclusion}

This paper has used Network Control Intelligence (NCI) to reinterpret the long pursuit of autonomous networking as a question of control capability, evidence, and authority. Across the eras reviewed here, progress did not follow a single ladder. It appeared instead as a sequence of partial advances: more programmable interfaces, richer operational state, stronger analytics, more expressive proposal mechanisms, and gradually broader but still bounded autonomy.

The central lesson is that network autonomy is not obtained by inserting a more intelligent component into an otherwise unchanged control stack. In credible designs, LLM-enabled components may generate proposals, explanations, candidate configuration changes, verification queries, or rollout plans, but operational changes must still be applied through established control and management systems. Authority is constrained by policy and invariant checks, authorization, staged rollout, rollback mechanisms, and change records. The enduring architectural question is not whether an agent can produce a plausible plan, but whether the surrounding system can justify committing it.

This perspective also clarifies what progress should mean. Larger models and more capable agents may improve interpretation and workflow preparation, but trustworthy autonomy will depend on control systems whose authority grows only with the evidence that supports it. Adaptation is central to this progression, yet it also creates risk: as models, policies, workflows, and network conditions change, the justification for stronger autonomy must be rechecked rather than assumed to persist. The path forward is incremental. Autonomy should expand only where actions are grounded in a faithful state, constrained by checkable decision logic, exposed through evidence-bearing interfaces, and recoverable when assumptions fail. The measure of success is not the appearance of end-to-end agency, but the ability to increase autonomy without losing safety, accountability, or operational control.

\section*{Acknowledgment}
The authors used Grammarly for typographical corrections, grammar checking, and language polishing. The authors reviewed and approved the final manuscript and remain responsible for all technical content, citations, and conclusions.

\ifCLASSOPTIONcaptionsoff
  \newpage
\fi


\begingroup
\let\url\nolinkurl
\bibliographystyle{IEEEtran}
\bibliography{references}
\endgroup
%



%


\begin{IEEEbiography}[{\includegraphics[width=1in,height=1.25in, clip,keepaspectratio]{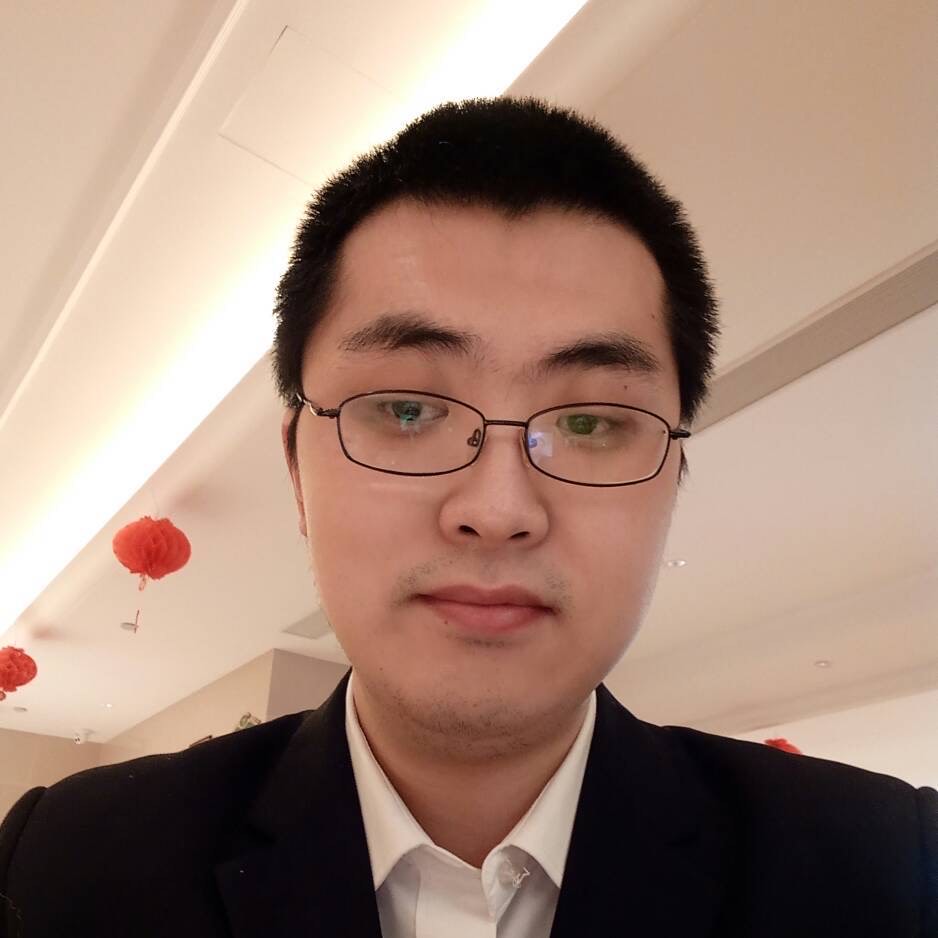}}]{Tianzhu Zhang}(Member, IEEE) received the B.S. degree in computer science and technology from Huazhong University of Science and Technology, Wuhan, China, in 2012, and the M.S. degree in computer and communication networks engineering from Politecnico di Torino, Turin, Italy, in 2014. From 2014 to 2017, he was a Ph.D. candidate in a joint program between Politecnico di Torino and Telecom Italia, supported by a Telecom Italia--PoliTo Ph.D. scholarship. From 2017 to 2019, he was a Postdoctoral Researcher at LINCS in France, supported by a research grant from Telecom ParisTech and Cisco Systems. He is currently a Research Scientist at Nokia Bell Labs and an Associate Member of LINCS. His research interests include software-defined networking, network function virtualization, high-speed network systems, AI/ML for networked systems, autonomous networking, and trustworthy network automation. 
\end{IEEEbiography}

\begin{IEEEbiography}[{\includegraphics[width=1in,height=1.25in, clip,keepaspectratio]{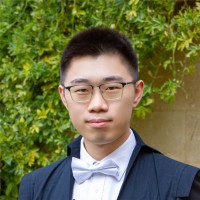}}]{Changgang Zheng} received the BEng degree in electronic and electrical engineering (first class Honours) from the University of Glasgow, Glasgow, U.K., and in communication engineering from the University of Electronic Science and Technology of China, Chengdu, China, and the DPhil degree in engineering science from the University of Oxford, Oxford, U.K. 
He was a Postdoctoral Researcher with the Computing Infrastructure Group at the University of Oxford in 2024 and conducted research at Alibaba Cloud in 2025. His research interests include data center networks, space networks, AI computing networks, in-network computing, programmable network devices, and machine-learning-based approaches for efficient and scalable network systems. His work has appeared in venues including NSDI, ACM CoNEXT, IEEE/ACM Transactions on Networking, ACM SIGCOMM Computer Communication Review, and IEEE Communications Surveys \& Tutorials.
\end{IEEEbiography}

\begin{IEEEbiography}[{\includegraphics[width=1in,height=1.25in,clip,keepaspectratio]{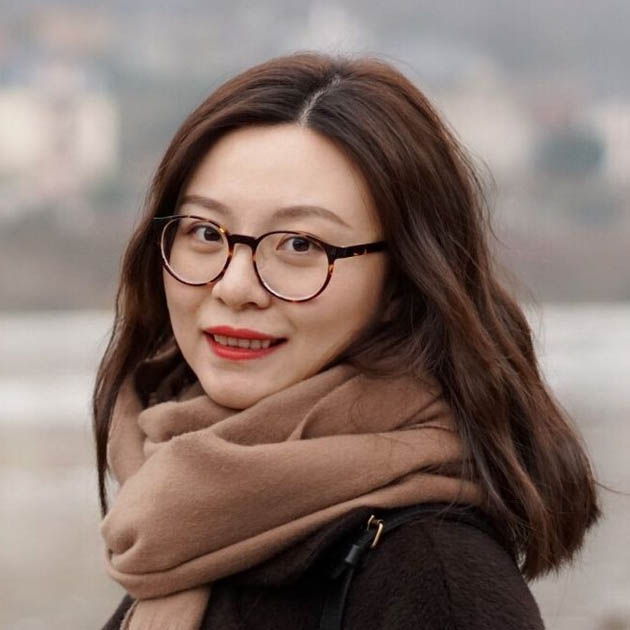}}]{Shanshan Wang} (Member, IEEE) received the master's degree with distinction from the University of Bristol, Bristol, U.K., in 2014, and the Ph.D. degree in modeling wireless networks from L2S, CNRS, Université Paris-Saclay, France, in 2019. She is currently an Assistant Professor with the Chaire C2M, Télécom Paris, Institut Polytechnique de Paris, and is affiliated with the RFM$^2$ team in COMELEC. Before joining Télécom Paris, she was a Research Engineer at the Toshiba Telecommunication Laboratory in Bristol, U.K. After receiving her Ph.D., she worked as a Postdoctoral Researcher at Télécom Paris's Chaire C2M, focusing on AI-based electromagnetic field exposure mapping. From 2023 to 2024, she was an Assistant Professor with ETIS, CY Cergy Paris University. She has participated in several European Horizon projects, including 5GWireless, SEAWave, and GOLIAT. Her research interests include electromagnetic-field exposure assessment, AI-based prediction, stochastic geometry, system-level modeling of wireless networks, machine learning, and uncertainty quantification.
\end{IEEEbiography}

\begin{IEEEbiography}[{\includegraphics[width=1in,height=1.25in,clip,keepaspectratio]{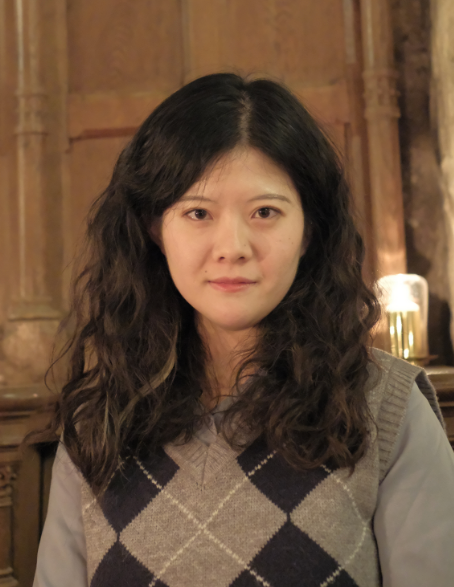}}]{Yarui Zhang} received the B.E. degree in electronic engineering from Xidian University, Xi'an, China, in 2018, and the M.S. and Ph.D. degrees in signal and image processing from Université Paris-Saclay, Gif-sur-Yvette, France, in 2019 and 2022, respectively. From 2023 to 2025, she was a Postdoctoral Researcher with the Chair C2M at Télécom Paris, France. She is currently an Assistant Professor with the SATIE Laboratory, École Normale Supérieure Paris-Saclay, France. Her research interests include inverse problems, computational imaging, and machine learning for wave-based problems.
\end{IEEEbiography}

\begin{IEEEbiography}[{\includegraphics[width=1in,height=1.25in,clip,keepaspectratio]{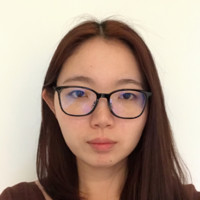}}]{Lina Shi}(Member, IEEE)  
received an engineering degree in Computer Science and Electronics for Embedded Systems from the Université Grenoble Alpes, Grenoble, France, in 2017, followed by a Ph.D. in telecommunication from Sorbonne Université, Paris, France. Between 2021 and 2022, she was a postdoctoral researcher at UVSQ (Université Paris-Saclay), France. She is currently a researcher at Nokia Bell Labs France. Her research interests include performance estimation and monitoring in optical transmission systems, machine learning, optical communication, and sensing.
\end{IEEEbiography}

\begin{IEEEbiography}[{\includegraphics[width=1in,height=1.25in,keepaspectratio]{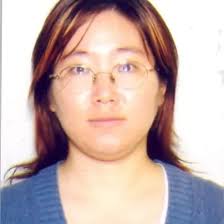}}]{Yue Jin}
received the B.S. degree in automotive engineering and the M.S. degree in industrial engineering from Tsinghua University, Beijing, China, in 2000 and 2002, respectively, and the Ph.D. degree in industrial engineering and operations research from the University of Massachusetts Amherst, Amherst, MA, USA, in 2007. She is currently a Researcher in the Machine Learning and Systems Department at Bell Labs France, Nokia. Her research interests include optimization, stochastic processes, reinforcement learning, and continual learning.
\end{IEEEbiography}

\begin{IEEEbiography}[{\includegraphics[width=1in,height=1.25in,clip,keepaspectratio]{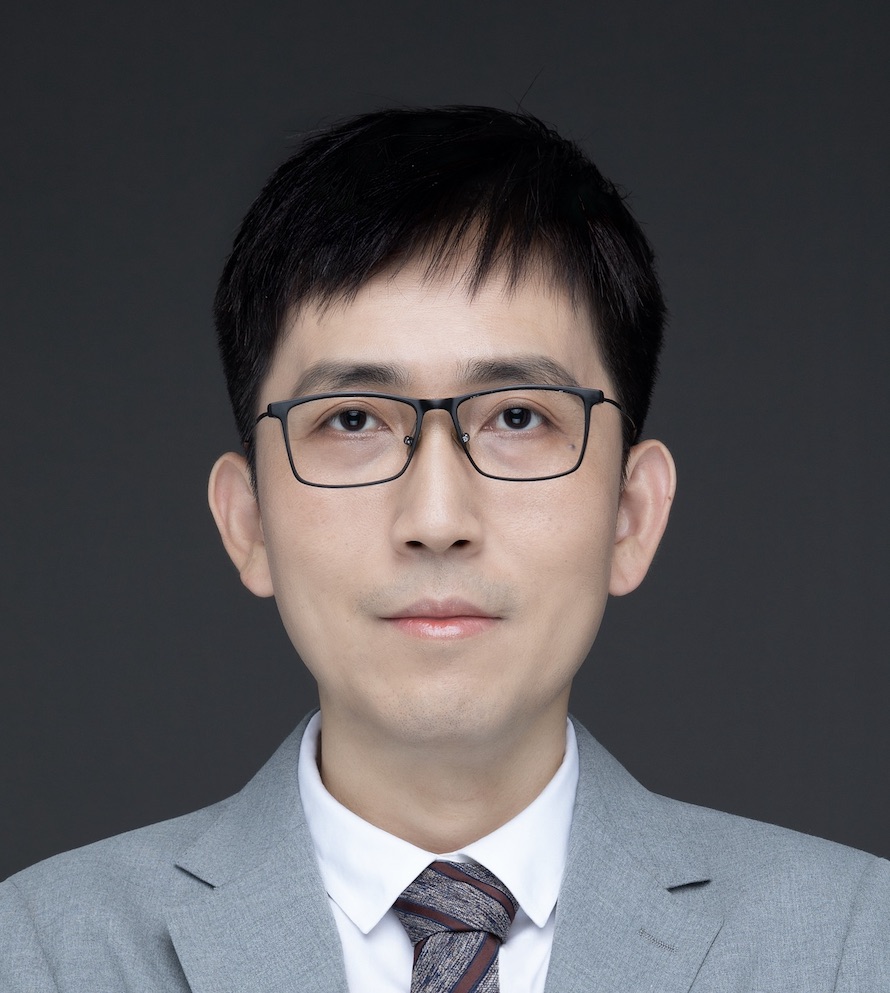}}]{Xiaofei Wang}
(Senior Member, IEEE) received the B.S. degree in computer science and technology from Huazhong University of Science and Technology, Wuhan, China, in 2005, and the M.S. and Ph.D. degrees in computer science and engineering from Seoul National University, Seoul, South Korea, in 2008 and 2013, respectively. From 2014 to 2016, he was a Postdoctoral Researcher with the School of Electrical and Computer Engineering, University of British Columbia, Vancouver, BC, Canada. He is currently a Professor with the School of Computer Science and Technology, Tianjin University, China. His research interests include edge computing, edge intelligence and cloud-edge collaborative systems. He has authored or coauthored more than 220 technical papers in journals and conferences, including IEEE Communications Surveys \& Tutorials, IEEE Journal on Selected Areas in Communications, IEEE Transactions on Mobile Computing, IEEE Transactions on Services Computing, IEEE/ACM Transactions on Networking, IEEE Transactions on Knowledge and Data Engineering, IEEE INFOCOM, and so on.
\end{IEEEbiography}

\begin{IEEEbiography}[{\includegraphics[width=1in,height=1.25in,keepaspectratio]{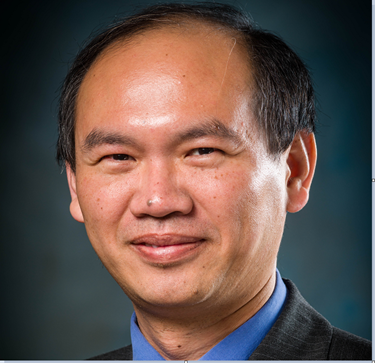}}]{Meikang Qiu}
(Senior Member, IEEE) received the B.E. and M.E. degrees from Shanghai Jiao Tong University, Shanghai, China, in 1992 and 1998, respectively, and the M.S. and Ph.D. degrees in computer science from the University of Texas at Dallas, Richardson, TX, USA, in 2003 and 2007, respectively. He is currently a Professor in the School of Computer and Cyber Sciences at Augusta University, Augusta, GA, USA. His research interests include artificial intelligence, cybersecurity, cloud computing, big data analytics, embedded systems, Internet of Things, bioinformatics, and smart computing. He has published extensively in these areas, including books, journal articles, and conference papers. He has served as an Associate Editor for several international journals, including IEEE Internet of Things Journal, IEEE Transactions on Computers, IEEE Transactions on Cloud Computing, IEEE Transactions on Big Data, and IEEE Transactions on Systems, Man, and Cybernetics: Systems. He was selected as a Highly Cited Researcher by Web of Science in 2020 and as an IEEE Distinguished Visitor from 2021 to 2023. He received the IEEE Bio-inspired Computing Special Technical Community Life-Career Award in 2023. He is an ACM Distinguished Member.
\end{IEEEbiography}






\end{document}